\documentclass[PAPER, cernpreprint, atlasdraft=false, texlive=2023, UKenglish, texmf, orcidlogo]{atlasdoc}
\usepackage{atlaspackage}
\usepackage{atlasbiblatex}

\usepackage{atlasphysics}
\graphicspath{{logos/}{figures/}{fig/}}

\usepackage{ANA-STDM-2018-60-PAPER-defs}

\AtlasTitle{Underlying-event studies with strange hadrons in $pp$ collisions at $\sqrt{s} = 13$ TeV with the ATLAS detector}

\AtlasAbstract{%
Properties of the underlying-event in $pp$ interactions are investigated primarily via the strange hadrons $K_{S}^{0}$, $\Lambda$ and $\bar\Lambda$, as reconstructed using the ATLAS detector at the LHC in minimum-bias $pp$ collision data at $\sqrt{s} = 13$ TeV. The hadrons are reconstructed via the identification of the displaced two-particle vertices corresponding to the decay modes $K_{S}^{0}\rightarrow\pi^+\pi^-$, $\Lambda\rightarrow\pi^-p$ and $\bar\Lambda\rightarrow\pi^+\bar{p}$. These are used in the construction of underlying-event observables in azimuthal regions computed relative to the leading charged-particle jet in the event. None of the  hadronisation and underlying-event physics models considered can describe the data over the full kinematic range considered. Events with a leading charged-particle jet in the range of $10 < p_T \leq 40$ GeV are studied using the number of prompt charged particles in the transverse region. The ratio $N(\Lambda + \bar\Lambda)/N(K_{S}^{0})$ as a function of the number of such charged particles varies only slightly over this range. This disagrees with the expectations of some of the considered Monte Carlo models.
}

\author{The ATLAS Collaboration}

\AtlasRefCode{STDM-2018-60}

\PreprintIdNumber{CERN-EP-2024-105}

\AtlasDate{9th May 2024}

\AtlasJournal{Eur. Phys. J. C}
\AtlasJournalRef{Eur. Phys. J. C 84 (2024) 1335}
\AtlasDOI{10.1140/epjc/s10052-024-13243-1}

\hypersetup{pdftitle={ATLAS document},pdfauthor={The ATLAS Collaboration}}

\begin{document}

\maketitle

\tableofcontents

\section{Introduction}

The simulation of proton-proton ($pp$) collisions at the Large Hadron Collider (LHC)
is divided into several stages. Typically there is the calculation of a hard interaction
in perturbative quantum chromodynamics (QCD), which is followed by QCD-inspired parton shower models and
then a phenomenological treatment of the hadronisation of the
resulting partons.
These hadronisation
models, such as the Lund string~\cite{Andersson:1983ia} and cluster~\cite{MARCHESINI1992465} models, attempt to act
as universal solutions: part of the description of almost all $pp$
collisions. The accuracy of these hadronisation models affects the interpretation of our measurements.

An extra complication is that partons from the $pp$ interaction, in addition to those in the hard scattering, can undergo scattering in a process known as multi-parton interactions (MPI). These additional
interactions are typically much softer than the hard interaction which, presumably, triggered the recording of the event.

Typical models of hadronisation have parameters that are tuned using data recorded in $e^+ e^-$ collisions from the LEP collider~\cite{OPAL:1998arz,OPAL:1995rqo}, as well as
from the LHC~\cite{STDM-2017-16} and other experiments. Much of these data uses
the properties of particles identified only as charged hadrons, but \Kshort and \LLbar
yields, and properties of strange mesons and baryons in general have also been important in determining these parameters. The parameter sets are known
as `tunes' when applied to models using the Monte Carlo (MC) method and are made public by the experimenters using them.
The increasing size of the high-precision LHC datasets demands ever higher
precision in the modelling, and motivates an expanded set of measurements to help tune the models. New approaches may also highlight regions where some models do not provide an adequate description.

The mass of the strange quark is close to the divergent scale of perturbative QCD, $\Lambda_{\textrm{QCD}} \sim 250~\MeV$, which makes for
a subtle interplay between kinematic effects and long-distance (low-energy) QCD interactions. The formation of baryons also presents
additional sensitivity to the modelling of MPI and colour reconnection via the three-way colour junctions required for baryon formation.

This paper presents spectra measurements within individual $pp$ interactions of the strange neutral hadrons \Kshort, \Lam and \Lbar, as measured in low pile-up $pp$ data recorded in 2015 by the ATLAS experiment at the LHC during \sqs = 13 \TeV\ collisions. These strange hadrons are reconstructed in this analysis via their displaced decay vertices, and these particle species are used extensively by LHC experiments to explore hadronisation and fragmentation in hadronic interactions; prior studies include Refs.~\cite{CMS-QCD-10-007, CMS-QCD-11-010, STDM-2010-09, CMS-HIN-11-011, ALICE:2019avo, ALICE:2021nvv, ALICE:2021vxl, ALICE:2020jsh, ALICE:2016fzo, ALICE:2020nkc, TOPQ-2016-05}.

The \LLbar and \Kshort yields are presented normalised to the total number of events or to the total number of prompt charged-particles, and the relative \LLbar to \Kshort yields are presented in addition. These ATLAS data are compared with MC predictions within a fiducial volume where ATLAS maintains both a high reconstruction efficiency and a low probability of fakes (a mis-reconstructed particle).

Measurements are made following a so called underlying-event (UE) formalism that originated in studies of $p\bar{p}$ collision in CDF~\cite{PhysRevD.70.072002}; many UE measurements are made based on LHC collision data with different particle species at different centre of mass energies, including ATLAS~\cite{STDM-2016-07}, ALICE~\cite{ALICE:2019mmy} and CMS~\cite{CMS-QCD-11-012}.

\begin{figure}
\centering
\includegraphics[width=.38\linewidth]{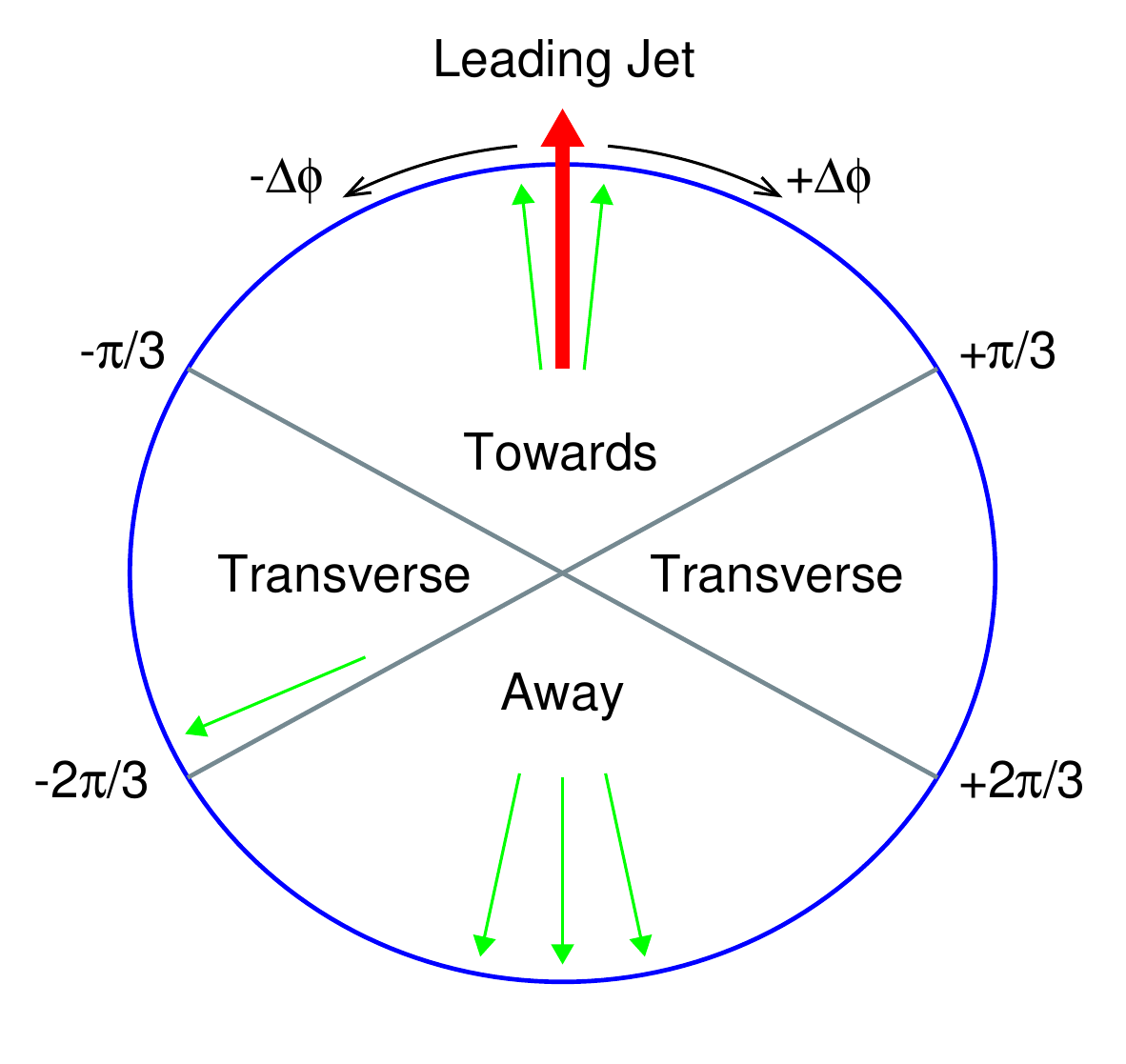}
\caption{Illustration of the underlying-event regions relative to the leading jet.}
\label{fig:UEDoodle}
\end{figure}

The charged-particle jet with the highest transverse momentum (\pt) in each event, denoted the `leading jet’, defines an axis in the plane transverse to the beam directions. The azimuth is split into three equal size regions defined by the leading jet as shown in Figure~\ref{fig:UEDoodle}. The towards region contains the $2\pi/3$ of the azimuth centred on the leading jet, the away region is $\pi$ away and typically contains most hadronic recoil. Finally the transverse region is formed from the two opposite-sided $\Delta\phi \simeq \pm\pi/2$ regions from the leading jet, each of size $\pi/3$. The transverse region is expected to be the most sensitive region in which to study hadronisation effects associated with MPI as it is minimally contaminated by any leading $2\rightarrow 2$ scattering process in the event.

The multiplicity of reconstructed \Kshort, of the combined sum of \Lam and \Lbar, and of prompt charged-particles is measured in each of the towards, transverse and away regions as a function of two variables. First, the multiplicities are measured against the \pt\ of the leading jet in the event. Second, they are measured against the multiplicity of prompt charged-particles in the transverse region $\left(\nchtrans\right)$. This variable exhibits sensitivity to per-event MPI fluctuations. This follows from the study performed in Ref.~\cite{Martin:2016igp} and complements other related measurements such as those performed by ALICE in Refs.~\cite{ALICE:2016fzo, ALICE:2019mmy, ALICE:2020nkc, ALICE:2023csm}.

These measurements test a set of fragmentation properties that ATLAS has not previously probed, allowing a range of predictions from different MC models to be compared with data. None of the models considered are entirely satisfactory over the observables, and these data could be used to improve future simulations. Other recent modelling improvements such as the baryonic colour reconnection model integrated into the \textsc{Herwig} MC generator in Refs.~\cite{Bellm:2019zci, Gieseke:2017clv} are also expected to be sensitive to the results in this paper.

The following chapters describe the ATLAS detector (Section~\ref{sec:detector}) and the data and simulation samples  (Section~\ref{sec:samples}). This is followed by the analysis' selections (Sections~\ref{sec:object_selections} and \ref{sec:event_selections}), correction factors (Section~\ref{sec:corr}), strategy (Section~\ref{sec:ana-strat}), uncertainties (Section~\ref{sec:err}), and ends with a set of results as compared with MC models (Section~\ref{sec:result}) and final conclusions (Section~\ref{sec:conclusion}).

\section{The ATLAS detector}
\label{sec:detector}

The ATLAS detector~\cite{PERF-2007-01} at the LHC covers nearly the entire solid angle around the collision point.\footnote{\AtlasCoordFootnote}
It consists of an inner tracking detector surrounded by a thin superconducting solenoid, electromagnetic and hadronic calorimeters,
and a muon spectrometer incorporating three large superconducting air-core toroidal magnets. The calorimeter and muon spectrometer detectors are
not used by the analysis reported in this paper.

The inner-detector system (ID) is immersed in a \SI{2}{\tesla} axial magnetic field
and provides charged-particle tracking in the range \(|\eta| < 2.5\).
The high-granularity silicon pixel detector covers the vertex region and typically provides four measurements per track,
the first hit generally being in the insertable B-layer (IBL) installed before Run~2~\cite{ATLAS-TDR-19,PIX-2018-001}.
It is followed by the SemiConductor Tracker (SCT), which usually provides eight measurements per track.
These silicon detectors are complemented by the transition radiation tracker (TRT),
which enables radially extended track reconstruction up to \(|\eta| = 2.0\).
The TRT also provides electron identification information
based on the fraction of hits (typically 30 in total) above a higher energy-deposit threshold corresponding to transition radiation.

The minimum-bias trigger scintillators (MBTS)~\cite{ATLAS-CONF-2010-068} provide the trigger signal.
These are mounted beyond the inner-detector volume at $z = \pm3.56$ m and are segmented into two rings in pseudorapidity
($2.07 < |\eta| < 2.76$ and $2.76 < |\eta| < 3.86$),
with eight azimuthal sectors in the inner ring and four in the outer.

Inelastic events are selected by the first-level trigger system implemented in custom hardware,
followed by selections made by algorithms implemented in software in the high-level trigger~\cite{TRIG-2016-01}.
The first-level trigger can accept events from the \SI{40}{\MHz} bunch crossings at a rate up to \SI{100}{\kHz}, but in the data sample used here it was pre-scaled to run at only about \SI{1.5}{\kHz} and all of the events were recorded without additional selection in the high-level trigger.

A software suite~\cite{SOFT-2022-02} is used in data simulation, in the reconstruction
and analysis of real and simulated data, in detector operations, and in the trigger and data acquisition systems of the experiment.

\section{Data and Monte Carlo samples}
\label{sec:samples}

Data from six LHC runs as recorded by the ATLAS detector in June 2015 are used. These runs had up to 29 colliding bunches with a large spatial separation between the colliding bunches, and with the probability of an inelastic interaction per bunch crossing being much smaller than one. Only data where the inner detector was operating nominally are included.

For five of the six runs,
events were predominately recorded by a single-hemisphere primary trigger that required that at least one MBTS sector was above threshold.
For a period during the sixth run, the primary trigger required at least one MBTS sector was above threshold in both the $+z$ and the $-z$ hemispheres of the detector.
This two-hemisphere trigger selection introduces a slight bias towards events with a larger number of charged particles over a wider range of pseudorapidity. Events recorded by the two-hemisphere trigger are only used in the second part of the
analysis that requires the event's leading jet to be in the range of \restrictedRange, as described in the following section.
There are about 110M events recorded with the single-hemisphere trigger and an additional 20M events recorded exclusively with the two-hemisphere trigger.

All triggered events are required to be in coincidence with time windows in which proton bunches were present and colliding in both of the beams in ATLAS\@.
The mean number of inelastic interactions per bunch-crossing, $\left<\mu\right>$, varied over the runs between $0.003 < \left<\mu\right> < 0.03$.

Three MC simulation samples are used: EPOS~\cite{Pierog:2009zt} using the EPOS-LHC~\cite{Pierog:2013ria} tune, \Pythia8 inelastic~\cite{Sjostrand:2014zea} with the A2~\cite{ATL-PHYS-PUB-2011-014} tune, and \Pythia8 inelastic with the Monash tune and modified colour-reconnection model~\cite{Christiansen:2015yqa} (which was used at particle level only). The ATLAS detector's response to the outputs of the EPOS-LHC and \Pythia8 A2 generators was simulated using \textsc{Geant4}~\cite{Agostinelli:2002hh,SOFT-2010-01}.

The EPOS MC provides an implementation of a parton-based Gribov-Regge theory~\cite{Drescher:2000ha}, which is an effective QCD-inspired field theory describing the hard and soft scattering simultaneously. The treatment of string hadronisation in EPOS is dependent on the local density of string segments per unit volume relative to a critical-density parameter. Each string is classified as being in either a low density coronal region or in a high-density core region. Corona hadronisation proceeds via unmodified string fragmentation whereas the core is subjected to a hydrodynamic evolution, i.e.\ it is hadronised including additional contributions from longitudinal and radial flow effects~\cite{Werner:2007bf}. This hydrodynamic collective flow approach to modelling MPI is contrasted against modelling in \Pythia8.

\Pythia8 inclusive $pp$ non-diffractive events are dominated by $t$-channel gluon exchange whereas diffractive interactions are modelled via colour-singlet exchange. The modelling in \Pythia8 is based on leading-logarithmic initial- and final-state parton showers, a Lund-string hadronisation model, and particle-decays and soft-QCD modelling, in contrast to the hydrodynamic collective flow approach of EPOS\@. As the partonic cross-section for the $t$-channel gluon exchange exceeds the hadronic non-diffractive cross-section at low \pt, the existence of MPI is implied and included in \Pythia8 as an eikonal distribution of perturbative QCD scatterings, governed by impact-parameter and hadronic-overlap functions. The ATLAS minimum-bias tune A2 is used. This tune is based on the MSTW2008 leading order (LO) parton distribution function (PDF) set~\cite{Martin:2009iq}, and was tuned using ATLAS minimum-bias data at 7 \TeV for the MPI parameters. It uses \Pythia's MPI-based colour reconnection scheme, with mergers of colour flow between MPI systems controlled by a reconnection-range parameter. It provides a good description of both the minimum-bias multiplicity and \pt distributions, and transverse energy-flow data~\cite{STDM-2011-44}.

The Monash~\cite{Skands:2014pea} tune is used in combination with an alternate colour-reconnection model to provide an additional simulation parameterisation which is compared with data. This is latter referred to as \Pythia8 Monash+CR (Colour Reconnection). The Monash tune was constructed using Drell–Yan and underlying-event data from ATLAS, but also data from CMS, from the Super Proton Synchrotron, and from the Tevatron in order to constrain energy scaling. It uses the NNPDF 2.3 LO PDF set~\cite{Ball:2012cx}. This tune gives an excellent description of the ATLAS 7, 8 and 13 \TeV minimum-bias \pT spectra~\cite{STDM-2015-02, STDM-2010-06, STDM-2014-19}. Monash is used in conjunction with the `Mode 2' parameterisation of the colour-reconnection model from Ref.~\cite{Christiansen:2015yqa}. This model uses approximations to the full group-theoretical weights from SU(3) to compute probabilistic string topologies where short string lengths are favoured. This allows `string-junction' structures to form, which provides a further source of baryon (and anti-baryon) production. This can arise at larger distances of order five femtometers, as compared with local baryon-production mechanisms.

\section{Object selections}
\label{sec:object_selections}

The following selections are imposed to identify candidate physics objects using the ATLAS inner-detector.

\paragraph{Prompt tracks:}
The prompt tracks selection only considers tracks reconstructed from the primary minimum bias tracking step, discussed in more detail in Ref.~\cite{STDM-2015-17}. This step reconstructs tracks with a maximum transverse impact parameter of 10 mm. Prompt tracks are required to satisfy $\pt > 500$~\MeV, $|\eta|~<~2.5$, minimum hits requirements in the pixel and SCT, and a $\chi^2$ requirement to remove mismeasured high-\pt tracks. Selected tracks are required to have both the transverse and longitudinal (multiplied by $\sin\theta$) absolute impact parameters relative to the primary vertex (see Section~\ref{sec:event_selections}) of less than 1.5 mm. Tracks with $|\eta| < 2.5$ are used as input to jet reconstruction, with a requirement of $|\eta| < 2.1$ being later used to construct underlying-event observables. This minimises contamination from tracks originating from jets with $|\eta| \geq 2.1$.

\paragraph{Jets:}
The \antikt algorithm~\cite{Cacciari:2008gp, Fastjet} with a radius parameter of \(R=0.4\) is used to reconstruct charged-particle jets using the set of prompt tracks as input. The leading jet is defined as the highest-\pt\ jet satisfying $|\eta| < 2.1$; this $\eta$ restriction prevents edge-effects by ensuring that tracks associated with the jet are all from within the tracking acceptance.

\paragraph{Large-radius tracks:}
ATLAS' standard tracking is not optimised for the reconstruction of charged particles at large impact parameter; the large-radius tracking runs as a secondary tracking step to improve upon the reconstruction efficiency for these particles. The reconstruction of these tracks is important to maintain efficiency for low-\pt charged particles from strange hadron decay which can have large curvature. This secondary step forms tracks with space points that were not used in the primary minimum bias tracking step. The transverse (longitudinal) impact-parameter requirements applied during reconstruction are loosened significantly from 10 to 300 mm (250 to 1,500 mm) and the requirement of a pixel hit is dropped. Tracks reconstructed in the large-radius step are not used directly; instead they are supplied as additional inputs to a \Vzero-finder algorithm.

\paragraph{\Vzero finder:}
Vertices reconstructed from a pair of oppositely-charged particle tracks are denoted `\Vzero'. The \Vzero-finder algorithm reconstructs candidate two-body decay vertices~\cite{STDM-2010-09}. The algorithm iterates over all possible pairs of oppositely charged particle tracks in the combined sample of both the primary and the large-radius tracks; no quality selections are applied to the tracks before their use in the \Vzero algorithm. The algorithm identifies two-particle vertex candidates with a $\chi^2$ probability $> 1\times 10^{-4}$. Candidates are rejected if the radius of the innermost space-point on either track is at a smaller radius than the \Vzero candidate reconstructed position. The passing \Vzero candidates form a preselection sample. They are fitted with each of the $\Kshort~\rightarrow~\pi^+~\pi^-$, $\Lam~\rightarrow~p\pi^-$ and $\Lbar~\rightarrow~\bar{p}\pi^+$ particle hypothesis for the positive and negative charged-particle track.

\paragraph{Kaon and Lambda:}
The selection of \Kshort, \Lam and \Lbar candidates is made from the set of preselected \Vzero. The selections favour a high-purity sample containing few fake \Vzero. The selections are listed in Table~\ref{tab:strange_sel}; here $\theta$ is the angle in 3D between the direction of the candidate momentum and the line joining the primary and \Vzero secondary vertices, \Rxy is the decay length of the candidate projected on to the $x,y$ plane, $M_{V^0}$ is the candidate's computed mass and $M^\textrm{err}_{V^{0}}$ is the uncertainty in this mass. Hadron masses are taken as $497.611\pm0.013$~\MeV\ for \Kshort and $1115.683\pm0.006$~\MeV\ for \Lam and \Lbar~\cite{ParticleDataGroup:2020ssz}.

\begin{table}[]
\centering
\caption{\Kshort, \Lam and \Lbar selection criteria.}
\label{tab:strange_sel}
\begin{tabular}{@{}lcc@{}}
\toprule
& \Kshort                               & \Lam, \Lbar                       \\ \midrule
$|\eta|$                  & $< 1.0$                                 & $< 1.0$                           \\
$\pt$                     & $> 400~\MeV$                            & $> 750~\MeV$                      \\
$\cos\theta$            & $> 0.9990$                              & $> 0.9998$                      \\
\Rxy                      & 4~mm$~< \Rxy \leq 300$~mm             & 15~mm$~< \Rxy \leq 300$~mm      \\
$M^\textrm{err}_{V^{0}}$ & $< 15~\MeV$                             & $< 5~\MeV$                        \\
$M_{V^0}$                 & $|M_{V^{0}} - M_{\Kshort}| < 20$~\MeV & $|M_{V^{0}} - M_{\Lam}| < 7$~\MeV \\ \bottomrule
\end{tabular}
\end{table}

Candidates must also satisfy two subsequent cleaning selections. Pairs of selected \Vzero of the same species are vetoed if they are less than a distance of $\Delta R < 0.1$ away from each other. This removes duplicate pairs of \Vzero that reconstruct from a triplet of tracks. A \Vzero is additionally vetoed if it simultaneously satisfies the \Kshort and \Lam, or \Kshort and \Lbar selections, which removes ambiguous candidates. It is not possible for a \Vzero to simultaneously satisfy the \Lam and \Lbar selections.

\begin{figure}
\centering
\includegraphics[width=.5\linewidth]{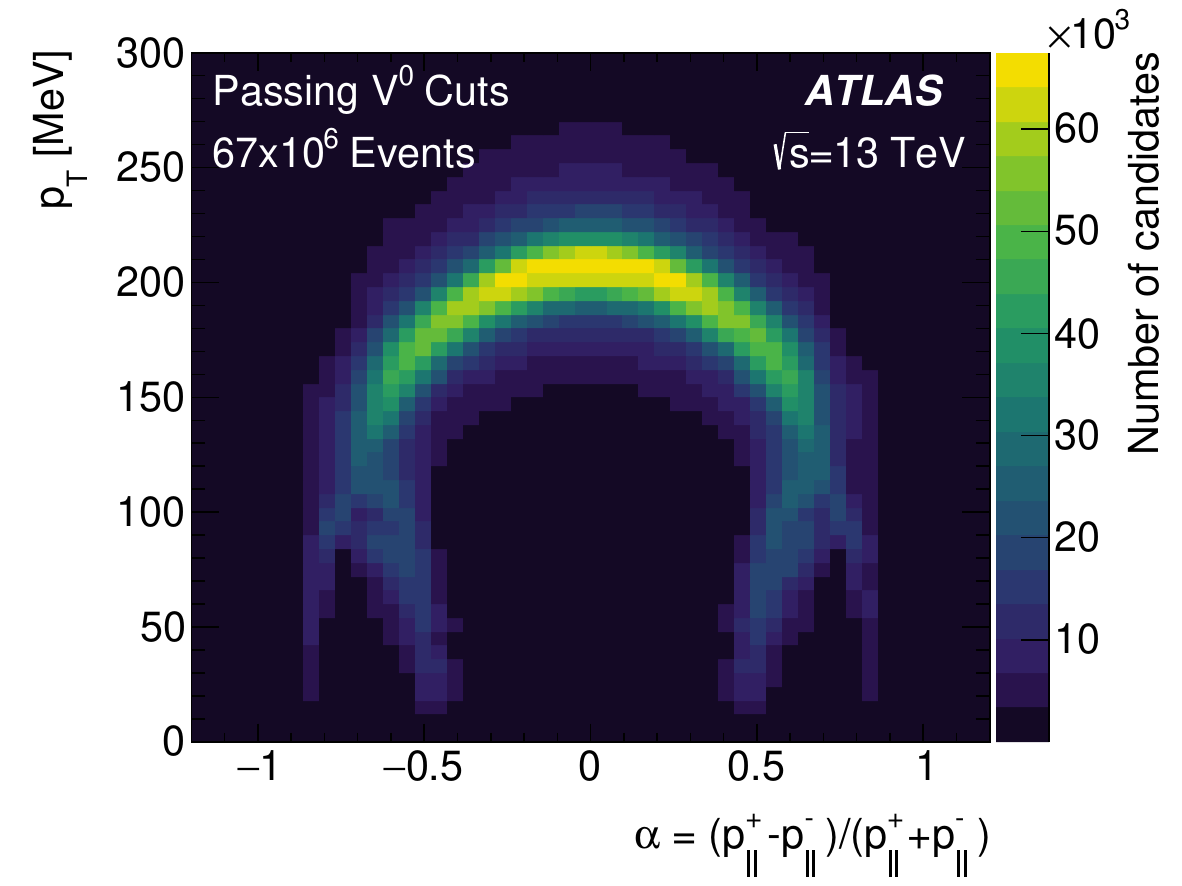}
\caption{Armenteros-Podolanski diagram of all \Kshort, \Lam and \Lbar candidates in data, the ordinate has bin intervals of 6 \MeV in \pt\ and the abscissa  has bin intervals of 0.048 in $\alpha$. See text for the definitions of the ordinate and abscissa.}
\label{fig:ap}
\end{figure}

Selected candidates are visualised in an Armenteros-Podolanski diagram~\cite{podolanski1954iii} in Figure~\ref{fig:ap}. In this 2D distribution the abscissa is used to plot the asymmetry of the momentum component of the \Vzero's two tracks that is parallel to the \Vzero's momentum vector, $\alpha = (p^+_\parallel-p^-_\parallel)/(p^+_\parallel+p^-_\parallel)$, and the ordinate to plot the momentum component of the two tracks that is perpendicular to the \Vzero's momentum vector. \Kshort have symmetric decays and hence a symmetric shape around $\alpha=0.0$ whereas \Lam populate the region near $\alpha=0.5$ and \Lbar the region near $\alpha=-0.5$ as the $p$ (or $\bar{p}$) takes a larger fraction of the decay momentum.

\paragraph{Particle-level selection:}
The particle level selection defines a fiducial volume, and the data are corrected and unfolded to match this definition. The particle level selection closely follows the reconstruction-level selection. Prompt particles are required to be stable (lifetime $\tau > 0.3\times 10^{-10}$~s), to not be a charged baryon with lifetime $0.3\times 10^{-10} < \tau\leq3.0\times 10^{-10}$ (see Ref.~\cite{STDM-2015-17}), $|\eta| < 2.1$ (or $|\eta| < 2.5$, when being used as input to jet finding), have a non-zero charge and $\pt > 500$~\MeV.

The reconstruction of charged-particle jets at the particle-level is analogous to the jets selection above at the reconstruction-level.

Branching ratios are folded into the selection of weakly decaying strange hadrons. In addition to the $|\eta|$, \pt\ and \Rxy requirements listed above in Table~\ref{tab:strange_sel}, selected \Kshort, \Lam and \Lbar are required to decay into $\pi^+\pi^-$, $p\pi^-$ and $\bar{p}\pi^+$, respectively. In all cases the two decay-children must satisfy $|\eta| < 2.5$, but there is no explicit \pt\ requirement on the children. Passing candidates of the same particle species that are less than a distance of $\Delta R < 0.1$ separation are vetoed to mirror the reconstruction-level selections.

\section{Event selections}
\label{sec:event_selections}

The following requirements are applied to select events for inclusion in the analysis. Physics objects as defined in Section~\ref{sec:object_selections} are used in the event selection.

\paragraph{Trigger:}
Data events are required to satisfy the trigger requirements detailed in Section~\ref{sec:samples}.

\paragraph{Vertex:}
Data and MC events are required to contain a primary vertex reconstructed from two or more tracks with $\pt > 100$~\MeV\ following the selection in Ref.~\cite{STDM-2015-17}.
Pile-up events are defined as data events containing two or more reconstructed primary vertices with four or more associated tracks each. These events are removed.

\paragraph{Track:}
Events are required to have at least one track with $\pt > 1$~\GeV\ which passes the prompt tracking selection. Particle-level events similarly require at least one selected charged particle with $\pt > 1$~\GeV.

\paragraph{Jet:}
Events where the leading jet satisfies \restrictedRange are classified as having satisfied the restricted leading-jet selection.

A total of 67M events satisfy the event selections in data, with a 1.4M event subset satisfying the restricted leading-jet selection. The available EPOS-LHC and \Pythia8 Monash+CR MC statistics are comparable in size to the data, whilst the \Pythia8 A2 sample is only around 25\% of the size of the data.

\section{Event and particle correction factors}
\label{sec:corr}

Some detector effects are corrected for by the application of correction factors. These are applied as an event-level weight for the trigger, and as per-candidate weights for selected tracks and \Vzero.

Data events selected by the single-hemisphere MBTS trigger are corrected to remove a small trigger inefficiency for low values of leading-jet \pt. This trigger is 98.8\% efficient in events with a single track reconstructed relative to the beam spot with $\pt > 500$~\MeV, and the efficiency quickly rises to 100\% with five or more tracks. No trigger correction is applied to the two-hemisphere MBTS triggered sample due to this only being used in conjunction with a \restrictedRange requirement on the leading jet, for which the trigger is 100\% efficient.

Selected prompt tracks and \Vzero are corrected for detector effects by application of a per-track or per-\Vzero weight. The initial computation of the weights with MC involves matching the selected prompt tracks and both of the decay tracks of the selected \Vzero to particle level tracks. Inner-detector space points are associated with the charged particle that was found to produce the largest energy deposition. Reconstructed tracks are matched to charged particles based on the fraction of space points on the track that are common to both the reconstructed track and the particle-level particle.

The MC-derived average tracking efficiency for prompt charged pions with $\pt > 500$~\MeV\ is 88\% at central pseudorapidity, falling to 75\% by $|\eta| = 2.1$. The track-correction weight includes efficiency corrections for charged particles that failed to be reconstructed, and corrections for reconstructed tracks that were not matched to particle level. The correction is assessed as a function of $\eta$ and \pt. Details of the correction are found in Ref.~\cite{STDM-2015-17}.

\Vzero are similarly corrected using MC-derived correction factors that are applied on a per-\Vzero basis; the details are as follows.

\begin{figure}
\centering
\subfloat[]{\includegraphics[width=.48\linewidth]{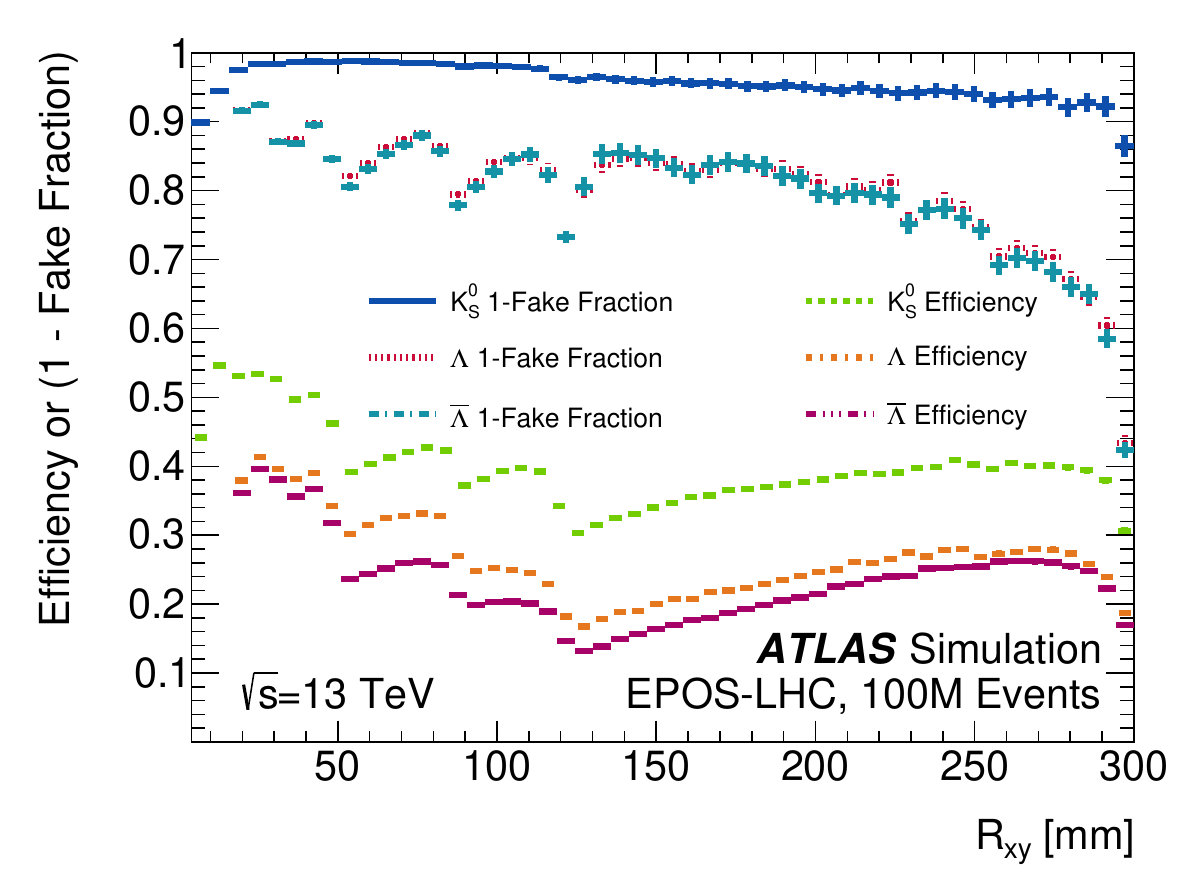}}
\subfloat[]{\includegraphics[width=.48\linewidth]{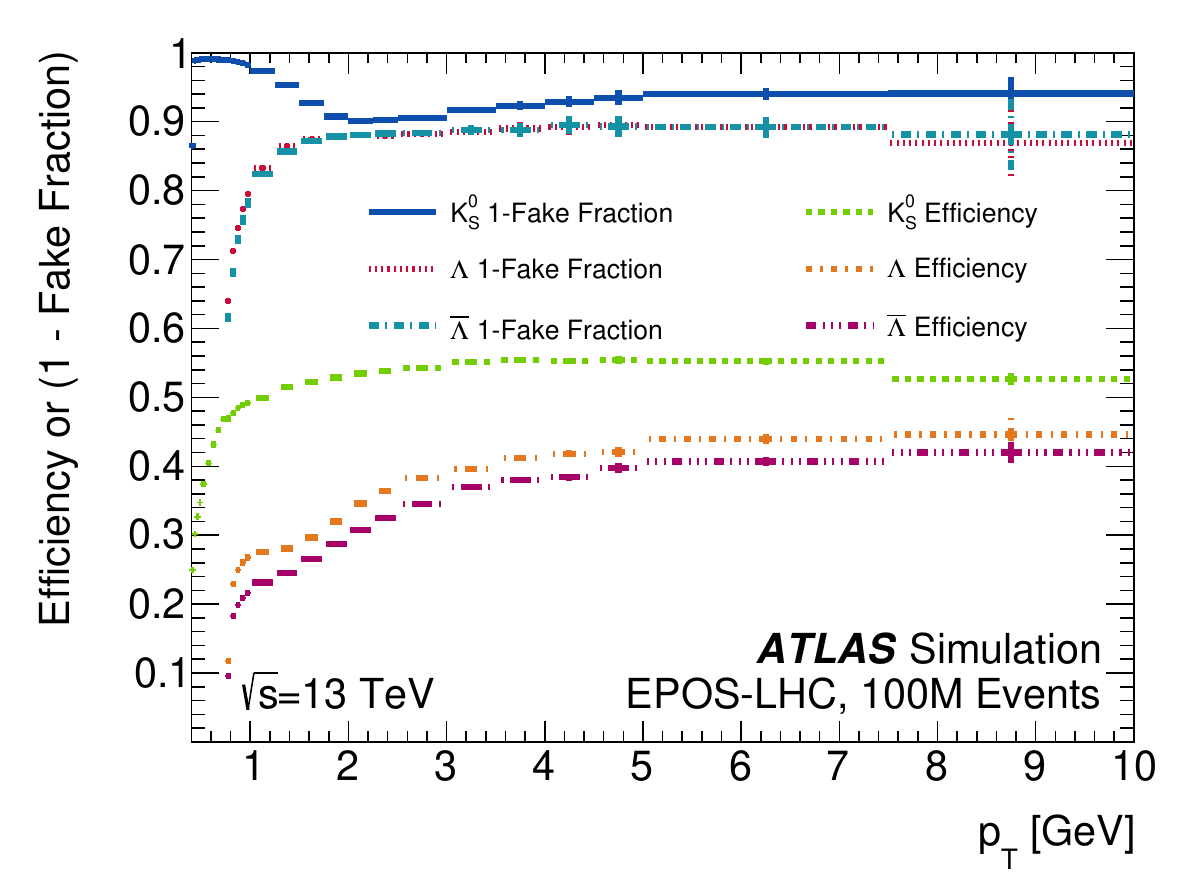}}
\newline
\subfloat[]{\includegraphics[width=.48\linewidth]{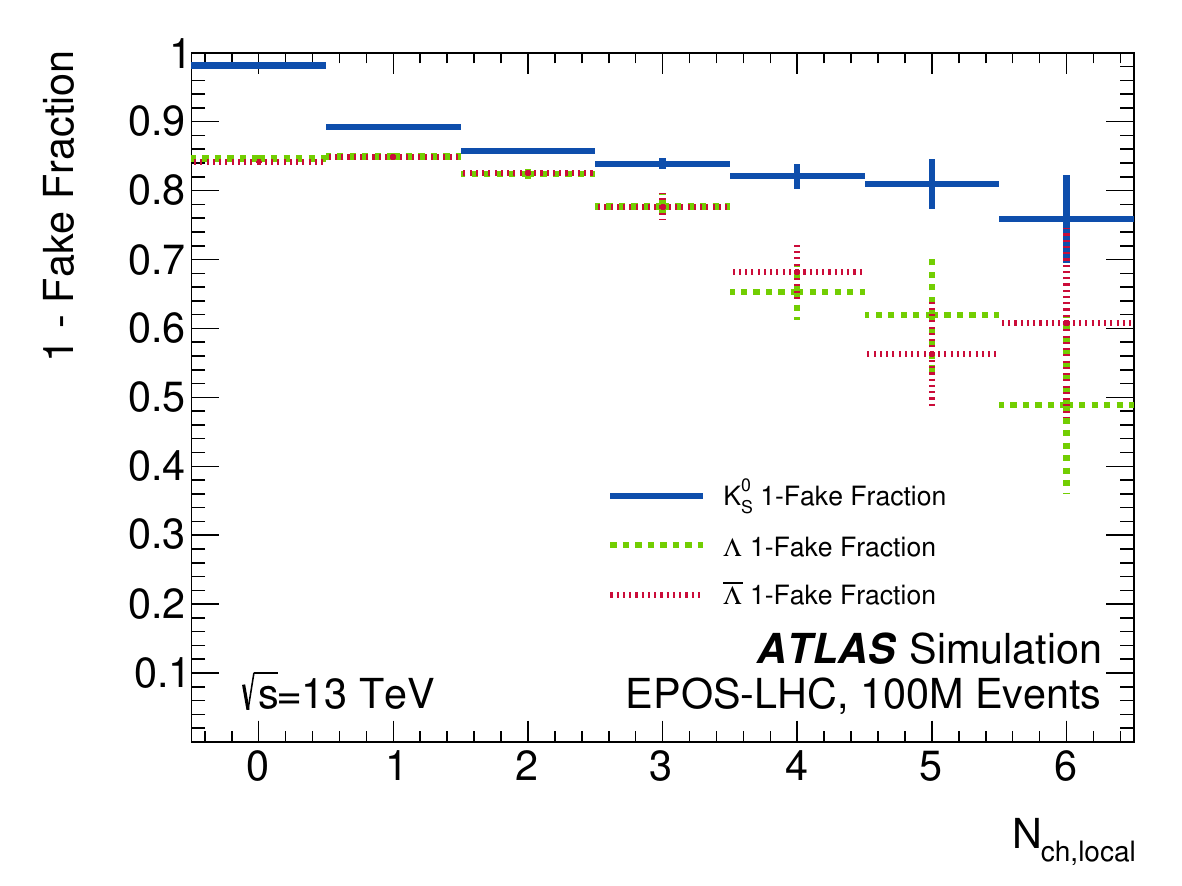}}
\caption{\Kshort, \Lam and \Lbar efficiency and ($1 - $fake fraction) as computed from EPOS-LHC MC for (a) \Rxy, (b) \pt, (c) \nlocalprompt. The structure as a function of \Rxy displayed in (a) is due to the effect of pixel detector layers.}
\label{fig:eff_fake}
\end{figure}

The \Vzero correction is factored into an efficiency correction, which is assessed as a function of $\pt$ and \Rxy, and a correction for fakes, which is quantified as a function of $\pt$, \Rxy and \nlocalprompt. Here \nlocalprompt\ counts the number of prompt selected tracks (i.e., from the primary vertex) within a cone of size $\Delta R = 0.2$ around the \Vzero's momentum vector. It allows the probability of a selected \Vzero arising due to combinatorial background or non-prompt strange-hadron production to be assessed as a function of the local charged-particle density. The probability that a \Lam candidate is a fake rises from 15\% for \nlocalprompt~=~0 up to around 50\% for \nlocalprompt~=~6. The \Vzero weight is the product of the efficiency weight ($w\!_\epsilon = 1/\epsilon, w\!_\epsilon \geq 1$) and the fakes weight ($w\!_f = 1 - \textrm{fake fraction}, w\!_f \leq 1$).
Examples of the \Kshort, \Lam and \Lbar efficiency and fake distributions are presented in Figure~\ref{fig:eff_fake}.

The mean correction factor is found by integrating over the fiducial region. The EPOS-LHC reconstruction efficiency is $46.07\pm0.01\%$ for \Kshort, $30.00\pm0.02\%$ for \Lam, and $26.20\pm0.02\%$ \Lbar. For comparison, the equivalent efficiency as computed with \Pythia8 A2 agrees to within $<0.01\%$ ($0.03\%$) for \Kshort (\Lam and \Lbar). The quoted uncertainties include only the statistical error. The difference between \Lam and \Lbar is due to the asymmetric momentum distribution in the decay together with the asymmetric efficiency of the ATLAS tracker relative to low-momentum charged-particles. The two MC simulations disagree in the fraction of selected \Vzero for which the MC particle record does not contain a strange hadron. This was found to arise from \Pythia8 A2 underestimating the inclusive \Lam and \Lbar yield by up to 50\% whereas EPOS-LHC is within 10\% of the data. Side-band studies were performed using the functional forms from Ref.~\cite{STDM-2010-09} to fit the \Kshort and \LLbar line-shapes. These studies indicate that data and both of the MC simulations have comparable levels of combinatorial background. The combination of the comparable background with an under predicted yield results in \Pythia8 A2 predicting that on average 30\% of the reconstructed \Lam and \Lbar candidates fail to match against a selected particle-level hadron, versus 25\% for EPOS-LHC\@. For \Kshort it is around 4\% for both of the MC simulations. EPOS-LHC is used to calculate the correction factors applied to the data due to its better agreement with data.

The strange-particle reconstruction efficiency was evaluated under systematically modified ATLAS geometry models derived from the studies in Ref.~\cite{ATL-PHYS-PUB-2015-050}. The largest effect was observed from a 5\% increase in all non-silicon material volumes in the inner detector. The mean efficiency under this geometry model was reduced by 0.7\% for \Kshort and 0.4\% for \LLbar, with the largest reduction for \Vzero at the low-\pt\ kinematic limit.

Systematic uncertainties are added for both the fake-fraction evaluation via side-bands and efficiency evaluation via material effects in simulation, as discussed in Section~\ref{sec:err}.

\section{Analysis strategy}
\label{sec:ana-strat}

For data and MC simulated events, the leading jet in selected events is used to define the towards, transverse, and away regions of the event. The multiplicity is computed separately in each of these three regions for the selected \Kshort candidates, the combined set of \LLbar selected \Vzero candidates, and for the selected prompt charged-particles. The per-candidate correction weights from Section~\ref{sec:corr} and Ref.~\cite{STDM-2015-17} are applied when computing these observables.
A further observable `event count' is added; this observable is used to count the events in each bin of (e.g.) leading-jet \pt and therefore allows per-event normalised figures to be formed. Final results are obtained by taking the ratio either between pairs of multiplicity observables, or as a multiplicity observable normalised to the event observable. This normalisation step is performed at the end of the analysis after all events are processed.

Each of the observable's distributions are unfolded before taking these ratios using an iterative method~\cite{DAgostini:1994fjx,Adye:2011gm} with four iterations, after which the process is observed to converge. Simulation from EPOS-LHC is used for the response matrix in the unfolding. The unfolding procedure corrects for migrations between different \pt bins of the reconstructed and particle-level leading jet. It does not correct for strange hadron or prompt charged-particle reconstruction effects; these were corrected for by the application of the per-candidate weights incorporated into the observables.

An initial set of results is presented as a function of leading-jet \pt. The \Kshort and \LLbar multiplicities are normalised to either the prompt charged-particle multiplicity in the same underlying-event region, or to the `event count' observable to obtain the mean yield.

This analysis strategy is repeated for events satisfying the restricted leading-jet selection, \restrictedRange. There is less of a dependency on the leading-jet \pt\ in this regime of higher-\pt\ hard-interactions, hence the choice of plot abscissa is changed to be the multiplicity of prompt charged-particles in the transverse region $\left(\nchtrans\right)$.

Ratios are presented as a function of \nchtrans for the \Kshort and \LLbar multiplicities normalised to the prompt charged-particle multiplicity in the towards region, and of the relative yield of \LLbar multiplicity to \Kshort multiplicity in the towards and transverse regions.

\section{Uncertainties}
\label{sec:err}

The following sources of systematic uncertainty are considered. They are assumed to be uncorrelated and are combined in quadrature. Unless noted, the systematic uncertainties are symmetrised around the central value. The unfolding prior, unfolding model, and non-closure systematic uncertainties below are all smoothed with a Gaussian kernel whose width is three times the width of the current bin, to mitigate the effect of statistical fluctuations.

\paragraph{Unfolding prior dependence:}
The dependence on the MC unfolding prior is evaluated using a reweighting method. The ratio is taken between MC simulation and data for each reconstructed observable and a spline-based interpolation between bins is used to convert this ratio into a weighting function. A reweighted variant of EPOS-LHC acting as pseudo-data is produced. For each event the reweighting function is evaluated based on the particle level leading-jet \pt or \nchtrans, the weight is applied to the observable at both the particle and reconstruction levels. The EPOS-LHC pseudo-data are unfolded through the nominal EPOS-LHC response matrix with the nominal number of iterations. The percentage deviation of the unfolded EPOS-LHC pseudo-data with respect of the reweighed EPOS-LHC particle level distribution is taken as a source of uncertainty.

\paragraph{Unfolding model:}
Data are unfolded through response matrices constructed using the \Pythia8 A2 MC sample as opposed to the nominal EPOS-LHC MC sample. The per-\Vzero correction factors applied to both the data and the \Pythia8 reconstructed \Vzero still wholly originate from EPOS-LHC\@. This procedure checks for differences between EPOS and \Pythia8 in the modelling of the leading charged-particle (required to be greater than 1~\GeV), the leading jet, and any differences between the modelling of event re-orientation (see below). It also checks against any residual effects from unfolding using a model with a different strange hadron yield, as \Pythia8 A2's \LLbar yield is significantly smaller than in data. The percentage difference between data unfolded with an EPOS-LHC response matrix and a \Pythia8 A2 response matrix is taken as a source of uncertainty.

\paragraph{Strange-hadron correction systematic uncertainties:}
A study was performed in which the probability of selecting a fake \Vzero was obtained by fitting the \Kshort, \Lam and \Lbar line-shapes as discussed in Section~\ref{sec:corr}. Inside of the analysis' mass-window, the contribution of the fit associated with combinatorial background differs by a maximum of 2\% between the EPOS-LHC MC simulation used to derive the correction factors and the data.

The effect of uncertainty in the material budget of the ATLAS detector model was also included in this systematic via a second study, as additionally discussed in Section~\ref{sec:corr}. The efficiency correction factor is modified by a power law that acts to reduce the efficiency by 4\% at a \Vzero \pt of 400~\MeV, and increase it by 2\%, clamped, for $\pt \geq 8~\GeV$.

Both the flat 2\% variation in the associated per-\Vzero fakes corrections weight and the power-law variation in the per-\Vzero efficiency corrections weight are varied independently and included in the strange hadron correction systematic for all of \Kshort, \Lam and \Lbar.

Final contributions in this category are derived from the statistical uncertainty in the evaluation of the efficiency and of the fake fraction. These two components of the \Vzero correction weight were independently varied positively and negatively by the respective asymmetric statistical uncertainty.

\paragraph{Prompt-track systematic uncertainties:}
Systematic uncertainties on prompt tracks from Ref.~\cite{STDM-2015-17} are included when the multiplicity of prompt tracks is evaluated. This includes the effects of inner-detector material uncertainty, the effect of the $\chi^2$ requirement applied to high-\pt tracks, plus an additional flat 0.5\% systematic.

\paragraph{Non-closure correction and systematic uncertainty:}
The EPOS-LHC reconstruction-level observables, including per-particle weighting, are unfolded through the EPOS-LHC response matrices. The term non-closure is used to refer to any residual differences that remain between a reconstruction level MC distribution that was fully corrected for all detector-effects and the corresponding particle level distribution from the same MC sample. Any non-closure is taken both as a correction factor, and as a conservative systematic uncertainty. The application of the correction is integrated into the bootstrap procedure. Many effects can cause this non-closure. Event re-orientation effects are where the $\phi$ of the leading reconstructed jet does not match the $\phi$ of the leading particle-level jet. This causes a misalignment of the three underlying-event regions between the particle level and the reconstruction level. The MC modelling of the effect is primarily encoded in the off-diagonal bins of the response matrices as it correlates with a mismatched jet \pt between particle level and reconstruction level. Other causes of non-closure effect arise from the \Vzero efficiency correction. This correction is computed in bins of particle-level strange hadron kinematics and is then applied on a statistical basis as a correction to the set of reconstructed \Vzero candidates.

\begin{figure}
\centering
\subfloat[]{\includegraphics[width=.48\linewidth]{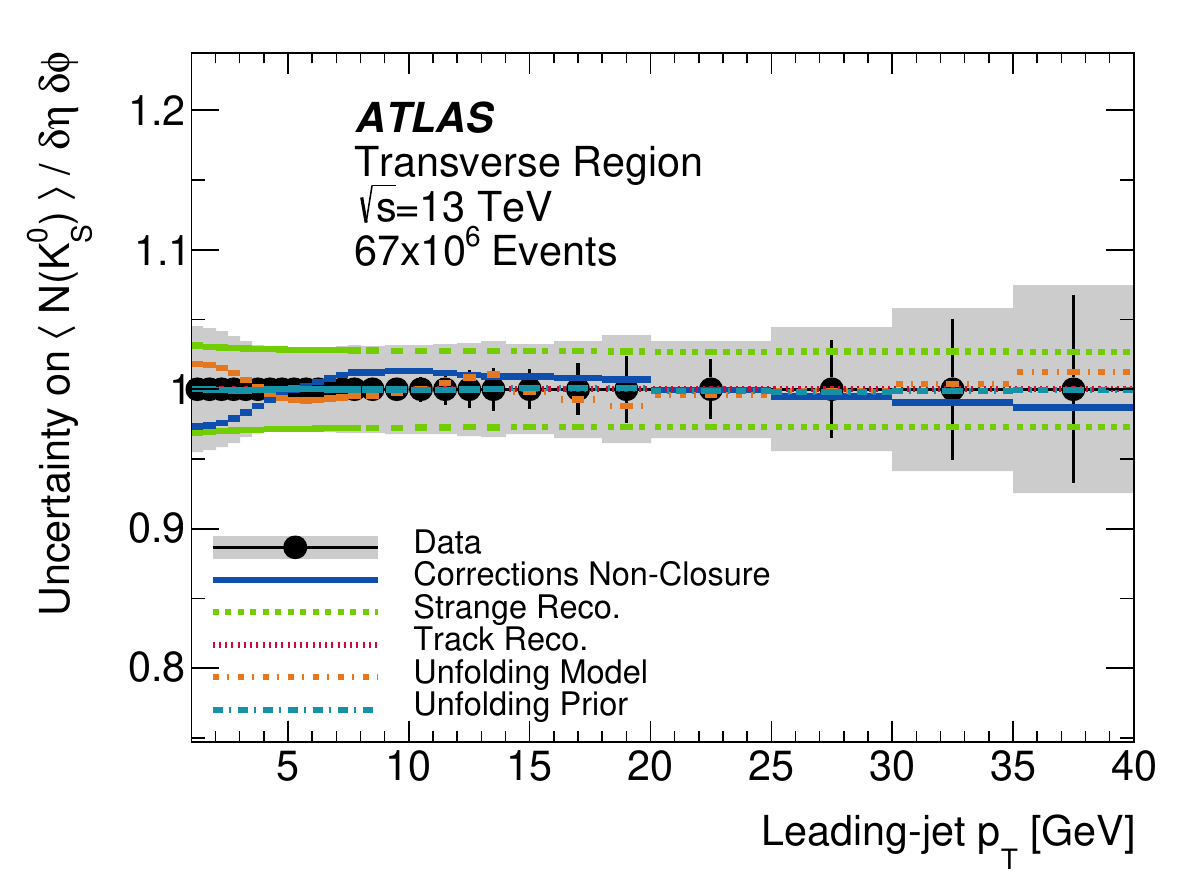}}
\subfloat[]{\includegraphics[width=.48\linewidth]{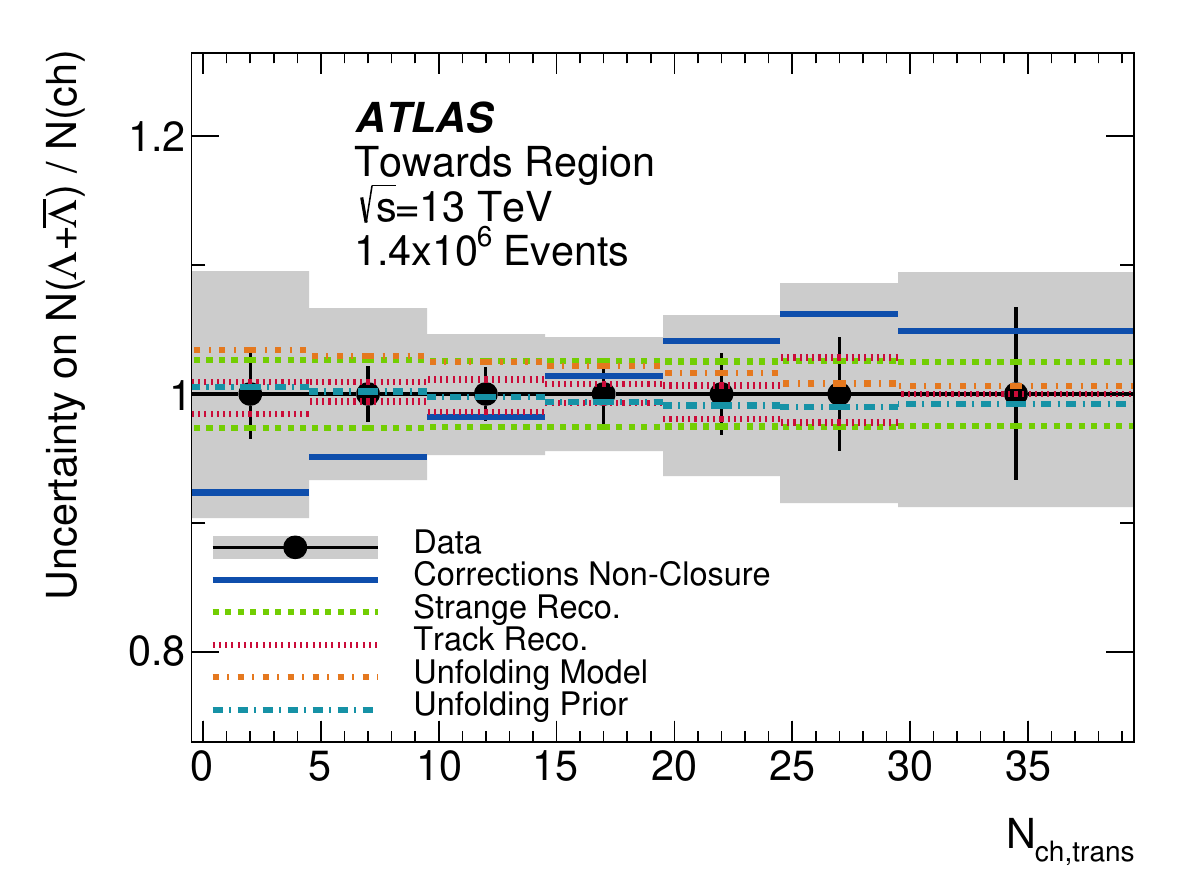}}
\caption{Breakdown of uncertainties in the measurement of (a) the event-normalised mean number of \Kshort in the transverse region as a function of leading-jet \pt, and (b) for the \LLbar multiplicity as normalised to the prompt charged-particle multiplicity in the towards region as a function of \nchtrans. Error bars show the statistical error and the shaded bands show the total uncertainty. The `Model', `Prior' and `Non-closure' systematic uncertainties are symmetrised.}
\label{fig:error_breakdown}
\end{figure}

Two illustrative breakdowns of the systematic uncertainties are shown in addition to the statistical uncertainty and the total uncertainty in Figure~\ref{fig:error_breakdown}, where one breakdown is chosen for each of the two choices of plot abscissa from the following results section. These systematic uncertainties are representative of the final Figures~\ref{fig:k_leadpt},~\ref{fig:l_leadpt} and \ref{fig:nch_ratio}. All of these figures are obtained from the ratio of two unfolded observables. A bootstrap method~\cite{ATL-PHYS-PUB-2021-011} is used to obtain the statistical error in each bin. This bootstrap technique uses 500 pseudo-runs in both the data and the MC simulation, each differing as expected across observables due to correlated statistical fluctuations. These correlations are propagated to all of Figures~\ref{fig:error_breakdown}, \ref{fig:k_leadpt}, \ref{fig:l_leadpt} and \ref{fig:nch_ratio}. The statistical error for each bin in all final figures is taken as the root mean square over all pseudo-runs; the covariance between pairs of bins is computed at the same time.

\section{Results}
\label{sec:result}
\begin{figure}
\centering
\subfloat[]{\includegraphics[width=.48\linewidth]{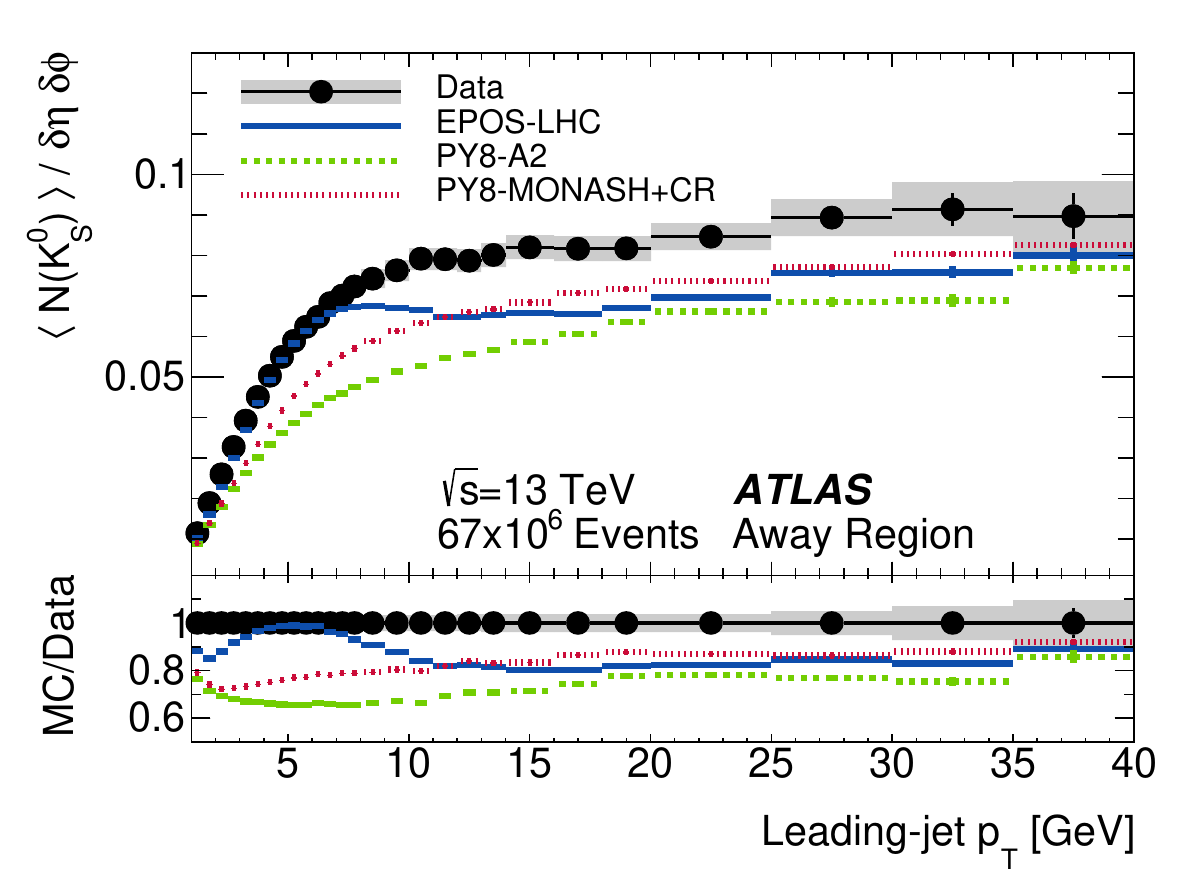}}
\subfloat[]{\includegraphics[width=.48\linewidth]{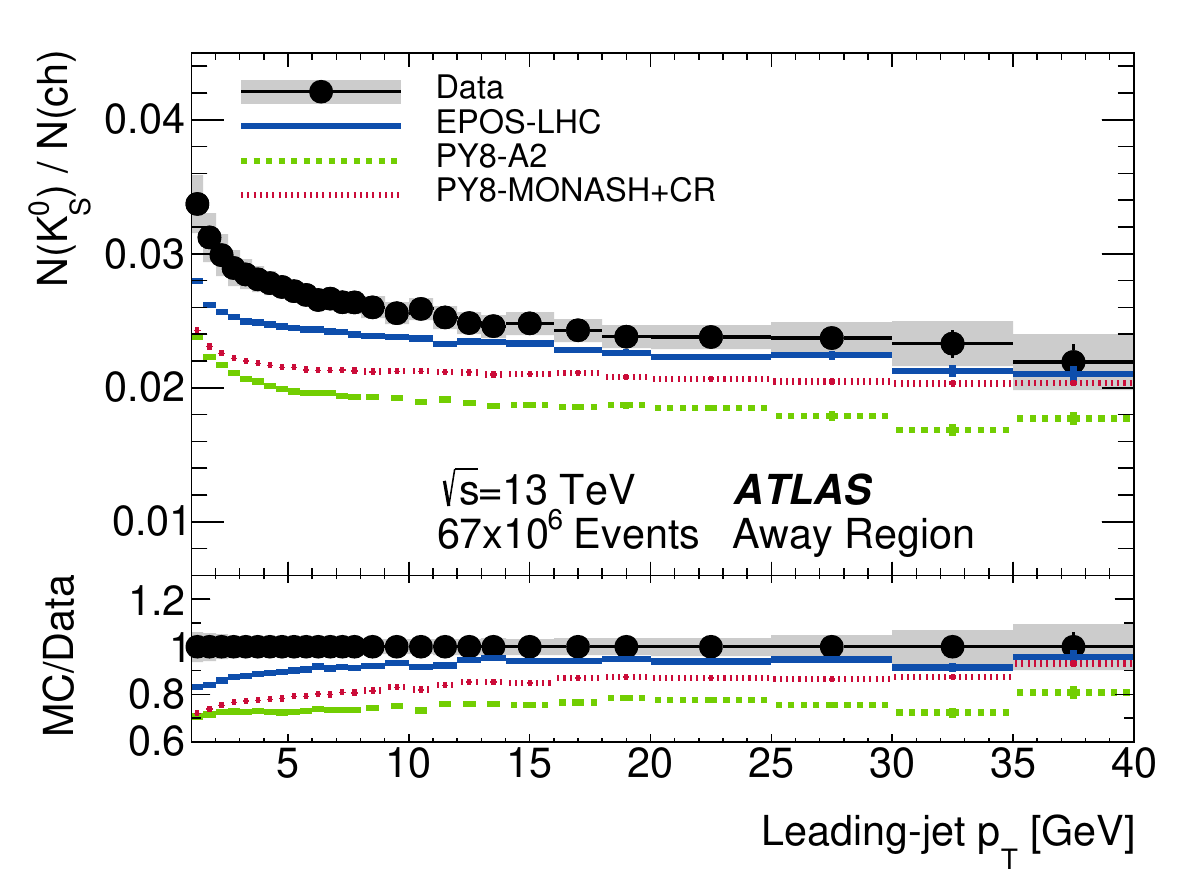}}\newline
\subfloat[]{\includegraphics[width=.48\linewidth]{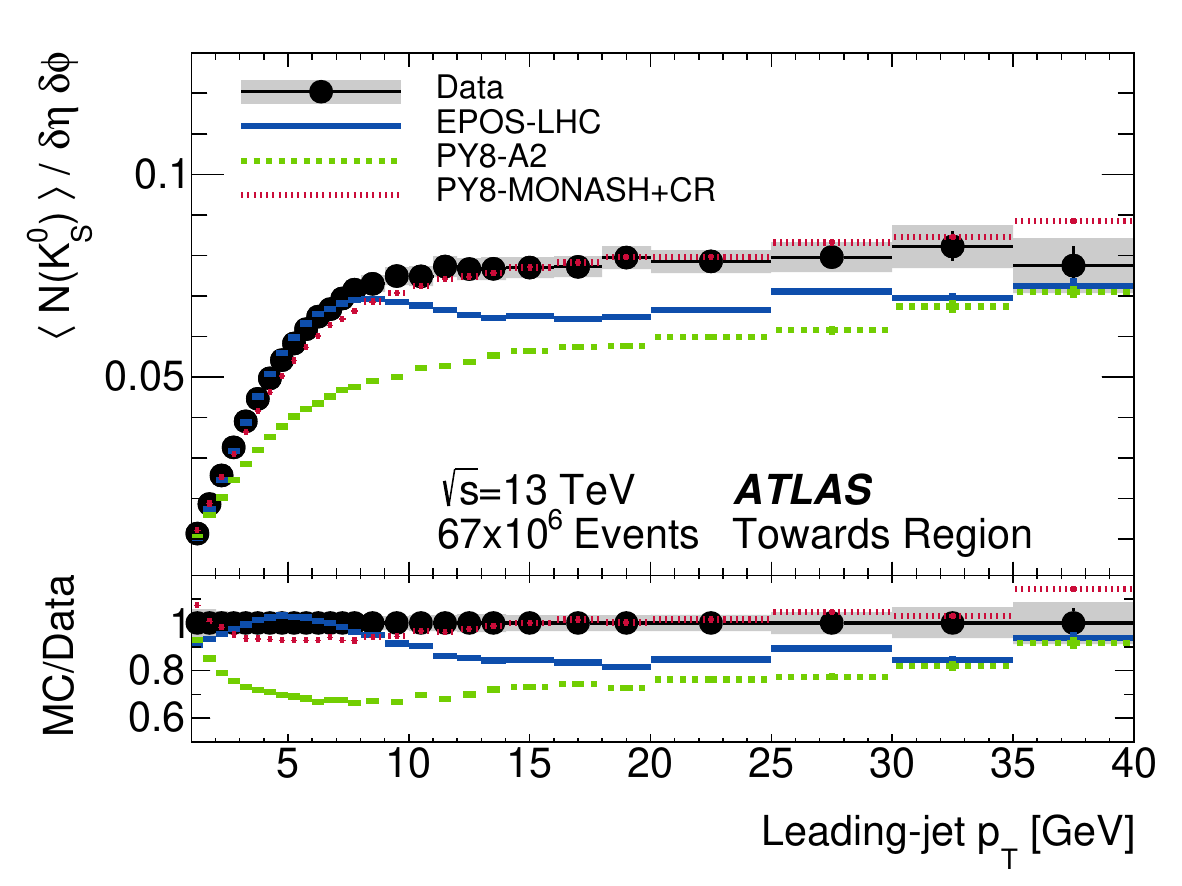}\label{fig:k_n_to}}
\subfloat[]{\includegraphics[width=.48\linewidth]{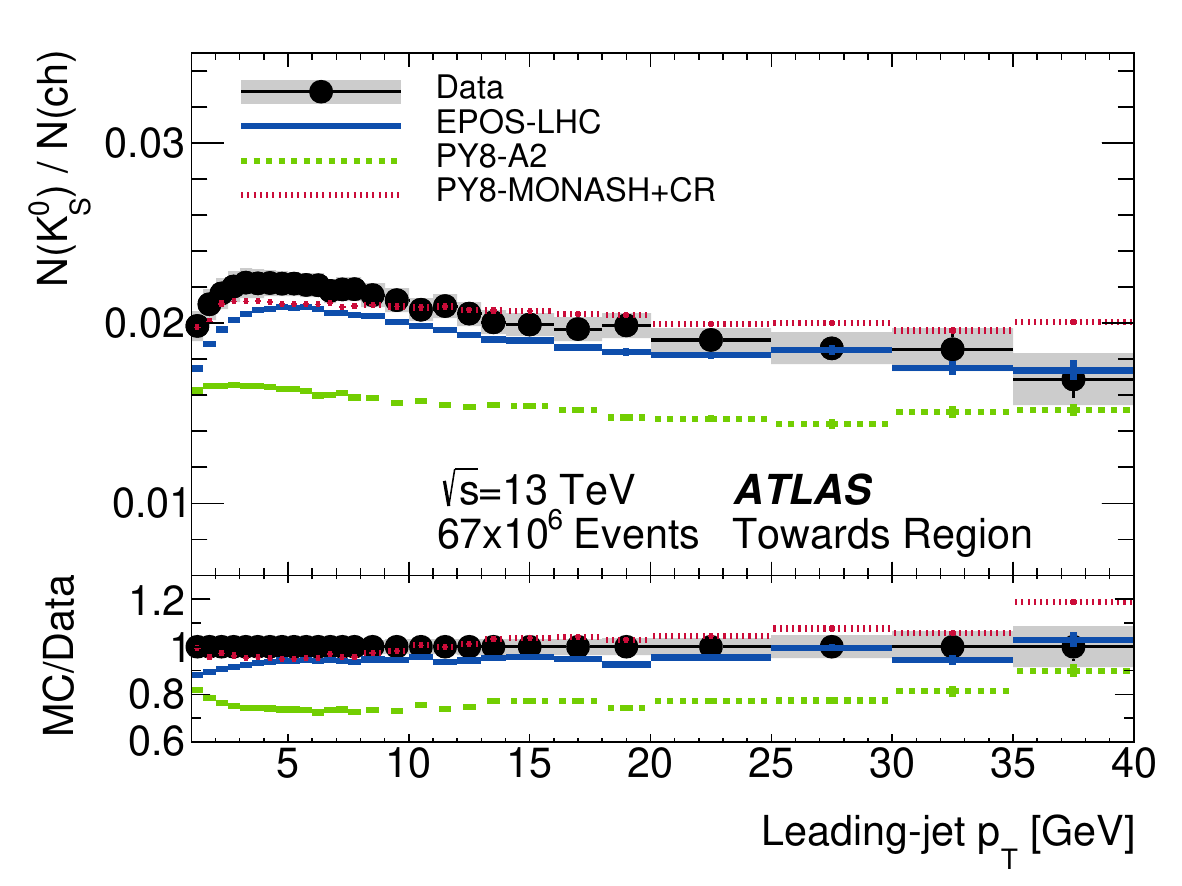}\label{fig:kc_n_to}}\newline
\subfloat[]{\includegraphics[width=.48\linewidth]{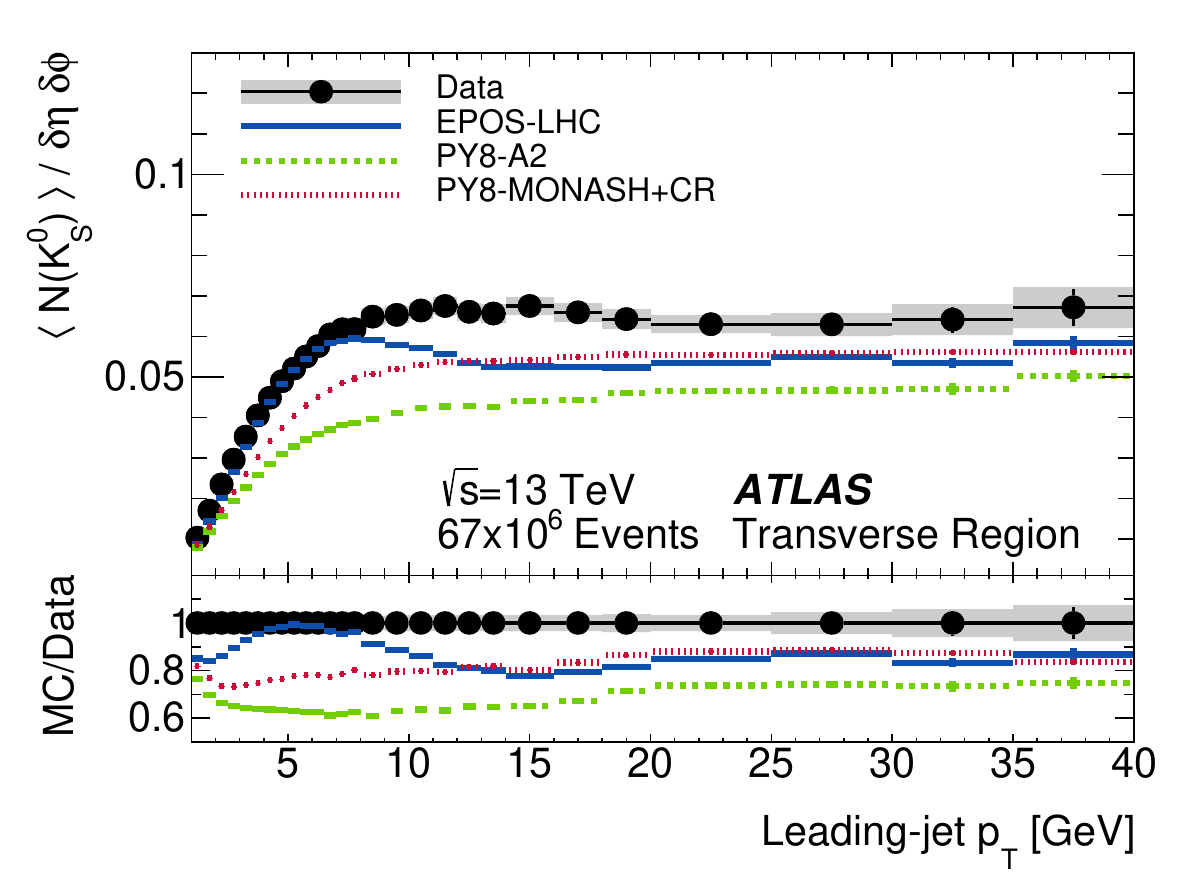}}
\subfloat[]{\includegraphics[width=.48\linewidth]{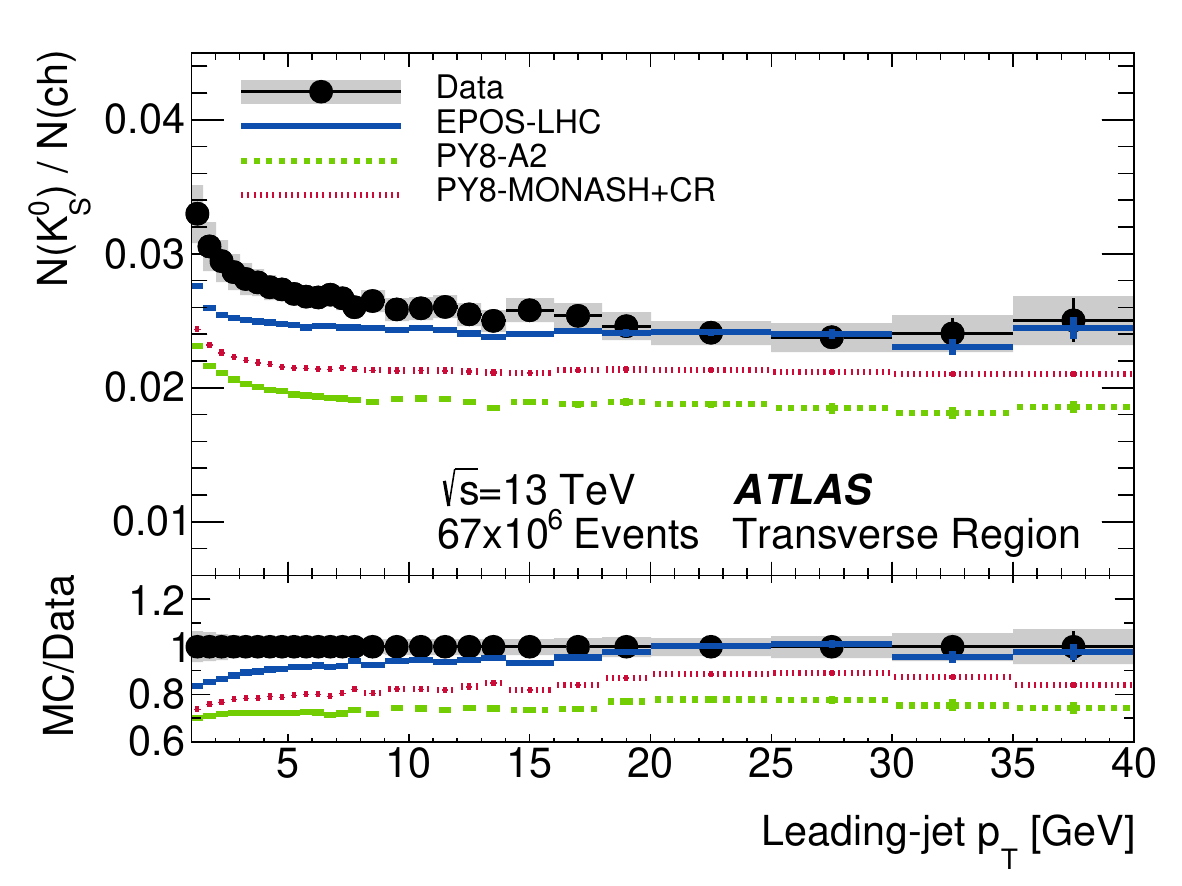}}
\caption{(Left) Per event and per unit $(\eta, \phi)$ normalised and (right) prompt charged-particle normalised \Kshort yields as a function of leading-jet \pt in the (a, b) away, (c, d) towards and (e, f) transverse regions. Error bars show the statistical error and the shaded bands show the combination of statistical and systematic uncertainties.}
\label{fig:k_leadpt}
\end{figure}

\begin{figure}
\centering
\subfloat[]{\includegraphics[width=.48\linewidth]{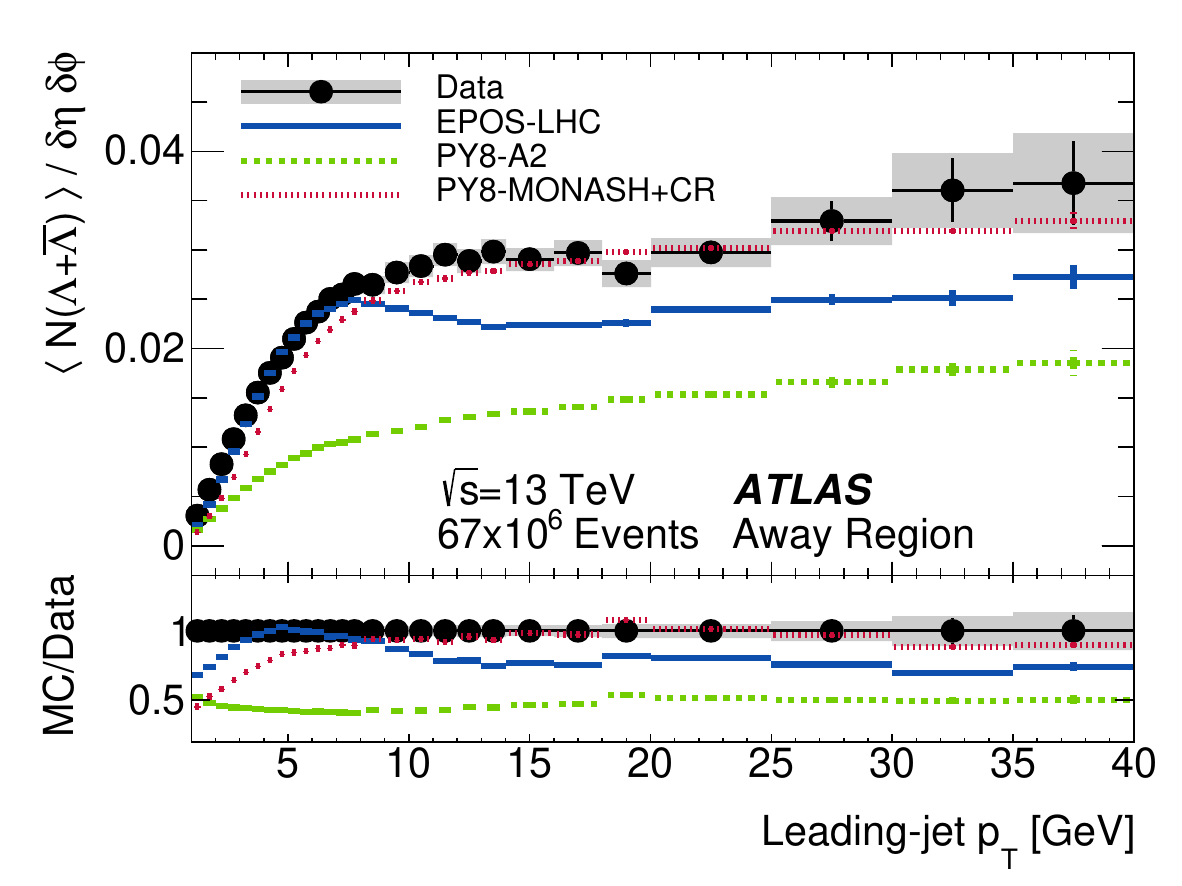}}
\subfloat[]{\includegraphics[width=.48\linewidth]{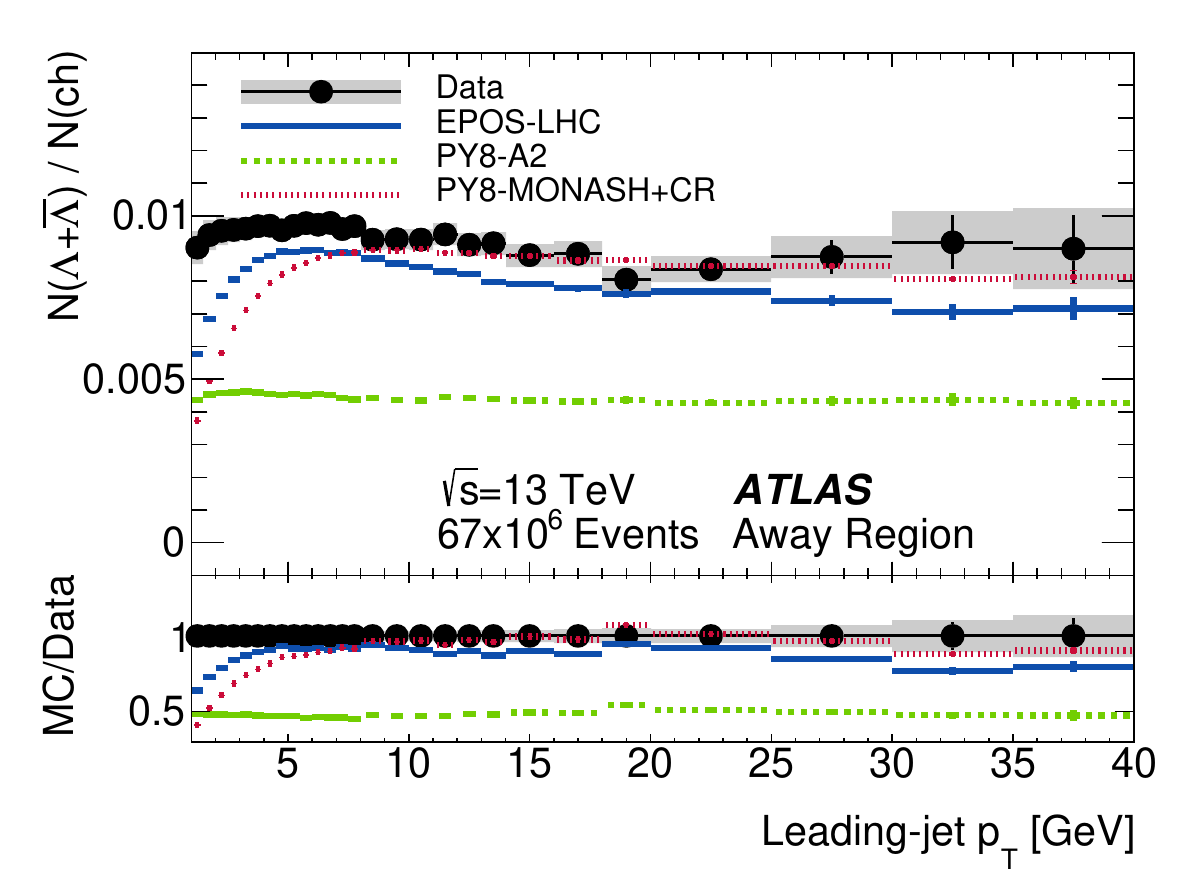}}
\newline
\subfloat[]{\includegraphics[width=.48\linewidth]{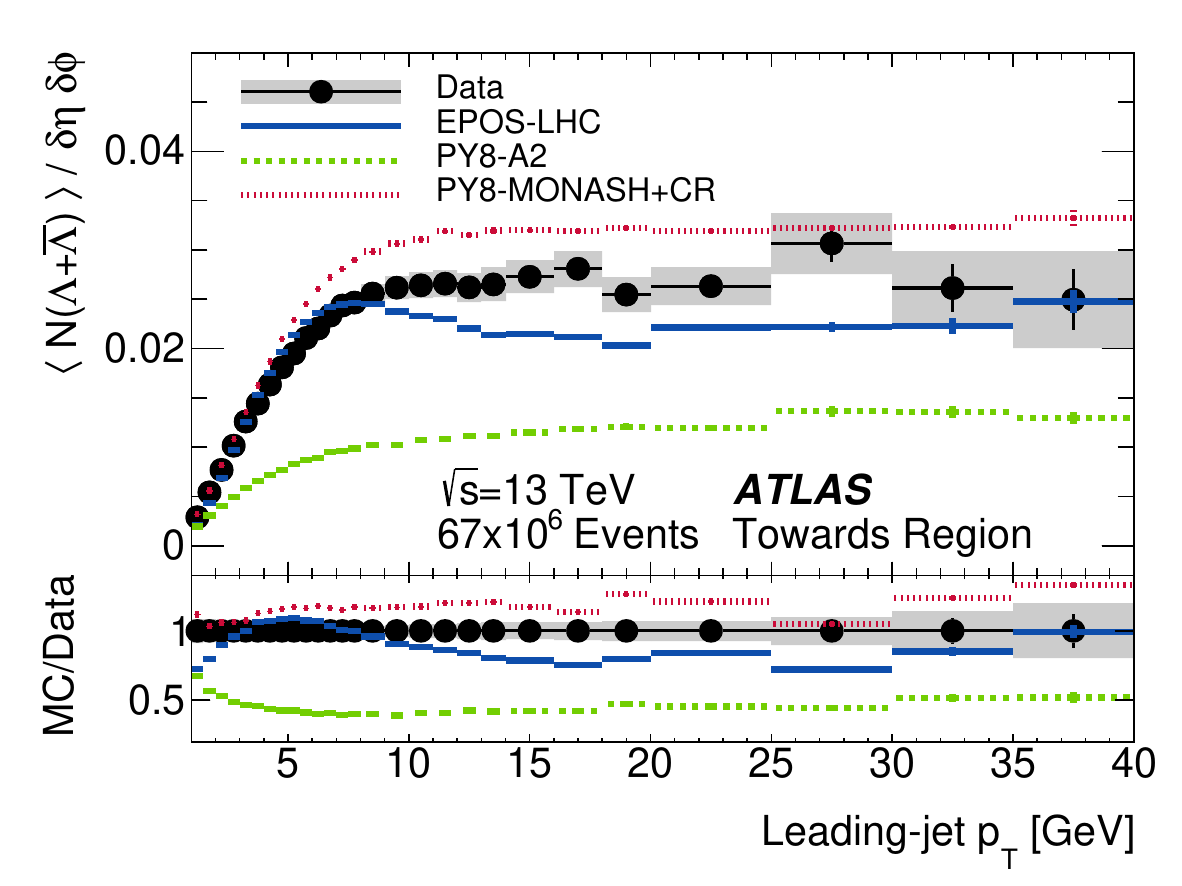}\label{fig:l_n_to}}
\subfloat[]{\includegraphics[width=.48\linewidth]{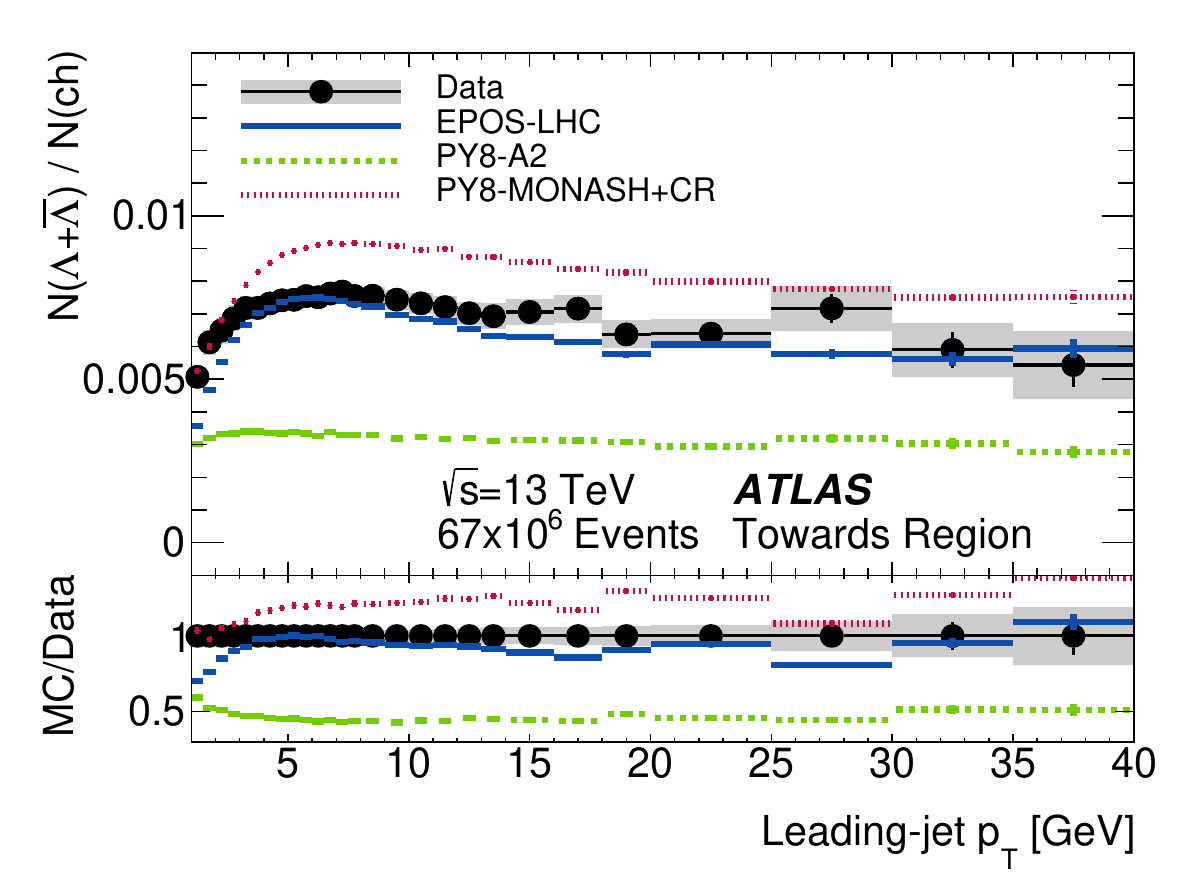}\label{fig:lc_n_to}}
\newline
\subfloat[]{\includegraphics[width=.48\linewidth]{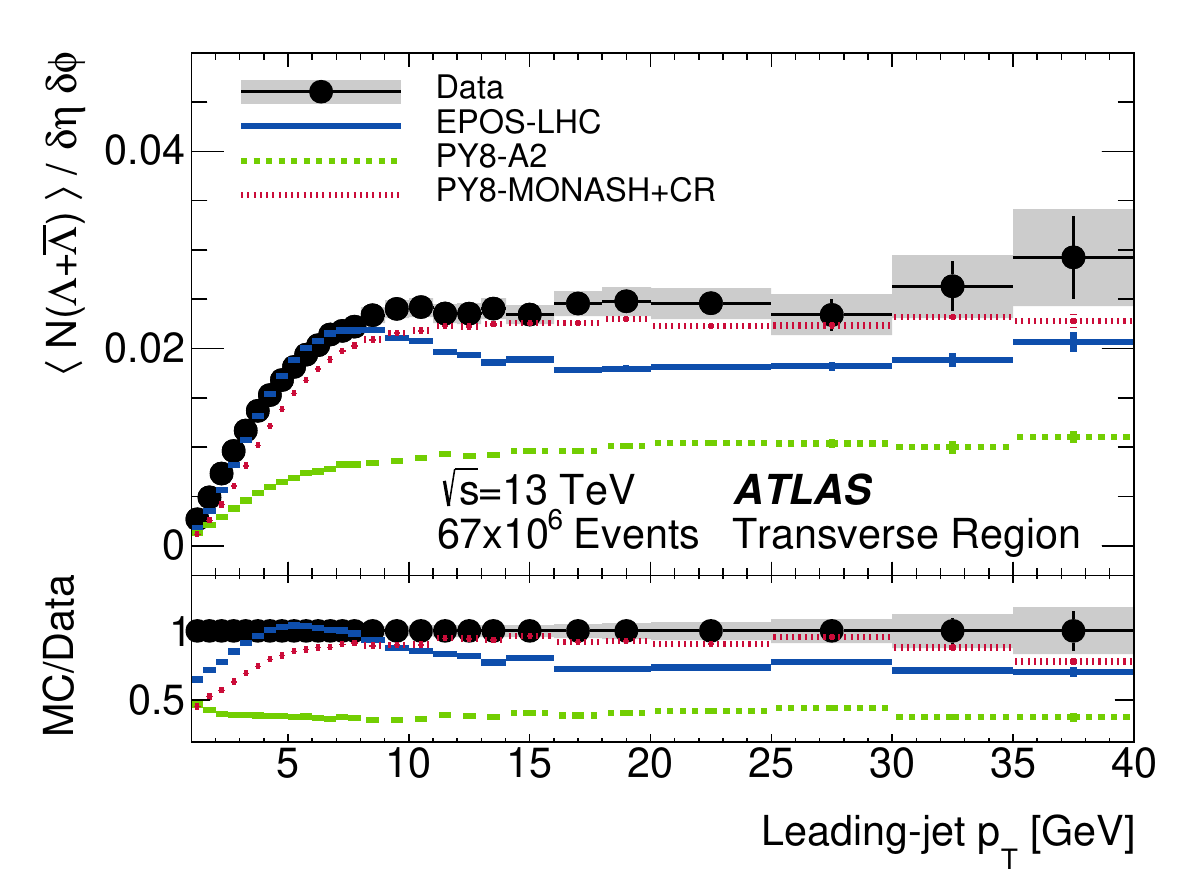}}
\subfloat[]{\includegraphics[width=.48\linewidth]{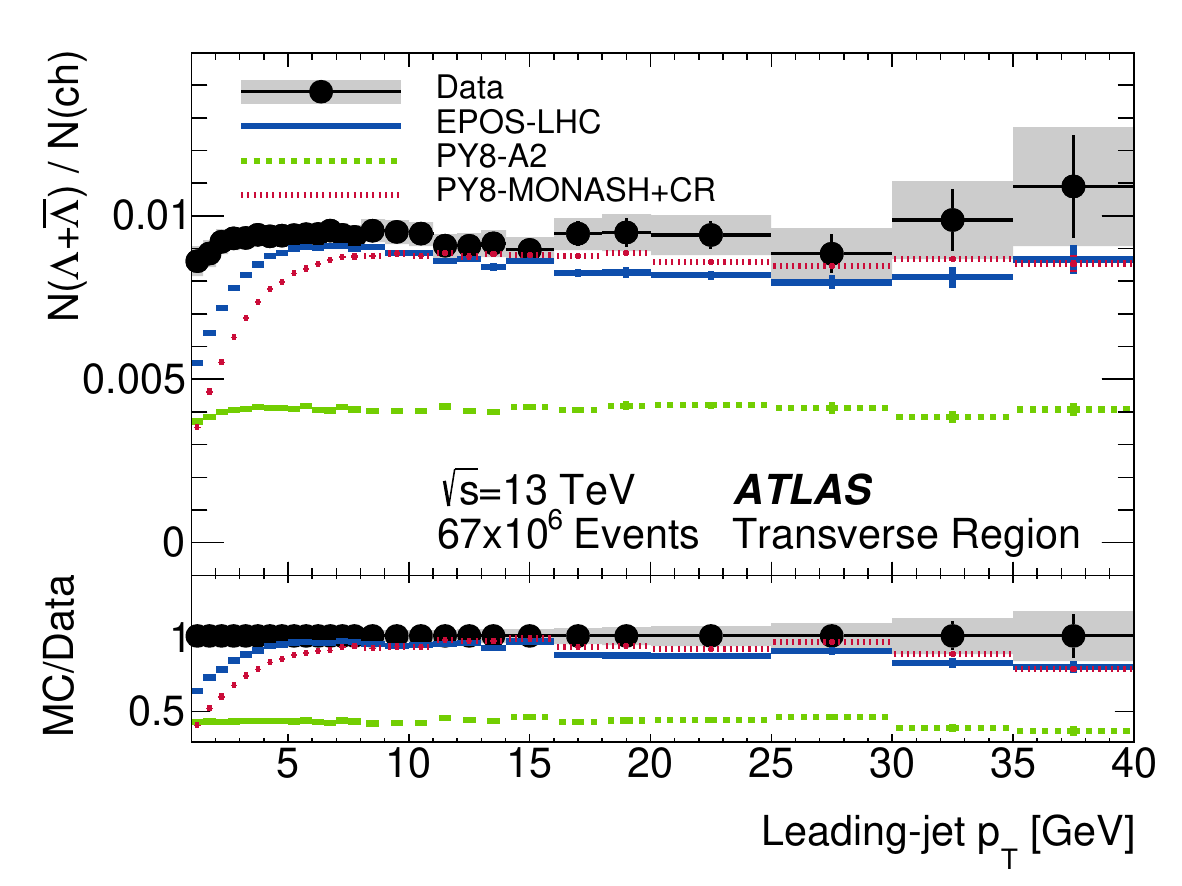}}
\caption{(Left) Per event and per unit $(\eta, \phi)$ normalised and (right) prompt charged-particle normalised \LLbar yields as a function of leading-jet \pt in the (a, b) away, (c, d) towards and (e, f) transverse regions. Error bars show the statistical error and the shaded bands show the combination of statistical and systematic uncertainties.}
\label{fig:l_leadpt}
\end{figure}

\Kshort and \LLbar multiplicities are presented in Figures~\ref{fig:k_leadpt} and \ref{fig:l_leadpt} as a function of the leading-jet \pt, with two choices of normalisation.
The event-normalised mean distributions show distinct soft and hard regimes, separated by a leading-jet \pt\ of around 10~\GeV.
The mean values in all distributions show a strong monotonic rise in the soft regime. In the hard regime the mean values either remain constant over the considered range, or continue to rise with a significantly smaller slope. This soft/hard transition is less distinct against the normalisation to the prompt charged-particle yield, here the strange-to-prompt yield is suppressed at low values of leading-jet \pt for \LLbar and for \Kshort in the towards region. However for \Kshort in the transverse and away regions it is enhanced. The soft regime is discussed first, followed by the hard regime.

No model gives a perfect description of the data, but the EPOS-LHC model gives the best overall performance in the soft regime for both of the choices of normalisation.
At a leading-jet \pt\ of around 5~\GeV, EPOS-LHC is observed to correctly model the normalisation for both of the \Kshort and \LLbar mean multiplicities over all underlying-event regions, the agreement is not as good when normalised to prompt charged-particles but EPOS-LHC remains in best or joint-best agreement. At very low leading-jet \pt, EPOS-LHC has a tendency to underestimate, especially for relative baryon yields. Similar and indeed larger trends are observed in the other models. An exception however is the Monash+CR model of the \Kshort and \LLbar relative yields in the towards region, as shown in Figures~\ref{fig:k_n_to}, \ref{fig:kc_n_to}, \ref{fig:l_n_to} and \ref{fig:lc_n_to}. The Monash+CR model generally performs as well or better than EPOS-LHC here in the towards region at low leading-jet \pt.

The \Pythia8 A2 model makes predictions that are  significantly too low in all distributions in the soft regime. The underestimation is large in magnitude,
up to 40\% for the \Kshort distributions in Figure~\ref{fig:k_leadpt} and up to 60\% for the \LLbar distributions in Figure~\ref{fig:l_leadpt}.
The \Pythia8 Monash+CR model performs better than the A2 model for all distributions.

The distributions show a much weaker dependence on the leading-jet \pt in the hard regime. Here the $pp$ interactions are predominately non-diffractive and at low impact parameter; the activity observed is hence driven by a combination of the multiple soft and semi-hard interactions occurring in each event that constitute the underlying-event and the activity from any hard scatter process. The towards and away regions maintain a tendency for the event-normalised mean multiplicity to rise as a function of increasing leading-jet \pt whereas it falls for the prompt-charged normalised cases. The slope is however considerably shallower in the hard regime for both of the choices of normalisation. The transverse region in data is predominately flat over the considered range of the hard regime between \restrictedRange.
This is in part by construction, as this region of the azimuth is minimally
affected by any leading partonic $2\rightarrow 2$ scattering interaction. It is noted that while the event-normalised mean is observed to have only a small
dependence on the leading-jet \pt, individual events may show large deviations from the mean, as explored below.

EPOS-LHC does not do as well in the hard regime as it did in the soft regime, likely a consequence of it lacking a hard-scattering model.
A common characteristic seen in all event-normalised EPOS-LHC distributions is a decrease in the mean multiplicity above approximately 8~\GeV\ in the leading-jet \pt, and a slower increase above 13 \GeV. This deviation from a monotonic rise over the transition from a soft to a hard event structure is not
observed in the other considered models, nor would such a dip hypothesis be drawn from the data. After this dip, the EPOS-LHC model predicts a slow monotonic rise in the event normalised yield and slow monotonic fall in the prompt charged-particle normalised yield,
the shape of which are generally in good agreement with data.
The overall normalisation following the dip region remains too low, however, in the event normalised sample when compared with data. EPOS-LHC under predicts the data by up to 20\% in the hard regime.

The \Pythia8 A2 model performs well in modelling the shape of the distributions in the hard regime. However, the large under predictions from the soft regime carry forward and the A2 model continues to under predict by 20\%--40\% for \Kshort distributions, and by 40\%--60\% for \LLbar distributions for both of the choices of normalisation. The shape of the \Pythia8 Monash+CR distribution is comparable to A2; however the Monash+CR model is shown to more accurately model an overall higher observed multiplicity in the hard regime. The \Pythia8 Monash+CR model is observed to be in good agreement with data when modelling the mean multiplicity of \Kshort in the toward region, and of \LLbar in both the away and transverse regions, for both of the choices of normalisation.

For a comparable UE measurement using only prompt charged particles such as Ref.~\cite{STDM-2016-07}, the agreement between data and MC simulation is typically 20\% or better, whereas for the strange-particle species in this analysis, differences between data and MC simulation exceeding 50\% are observed for some distributions.

\begin{figure}
\centering
\subfloat[]{\includegraphics[width=.48\linewidth]{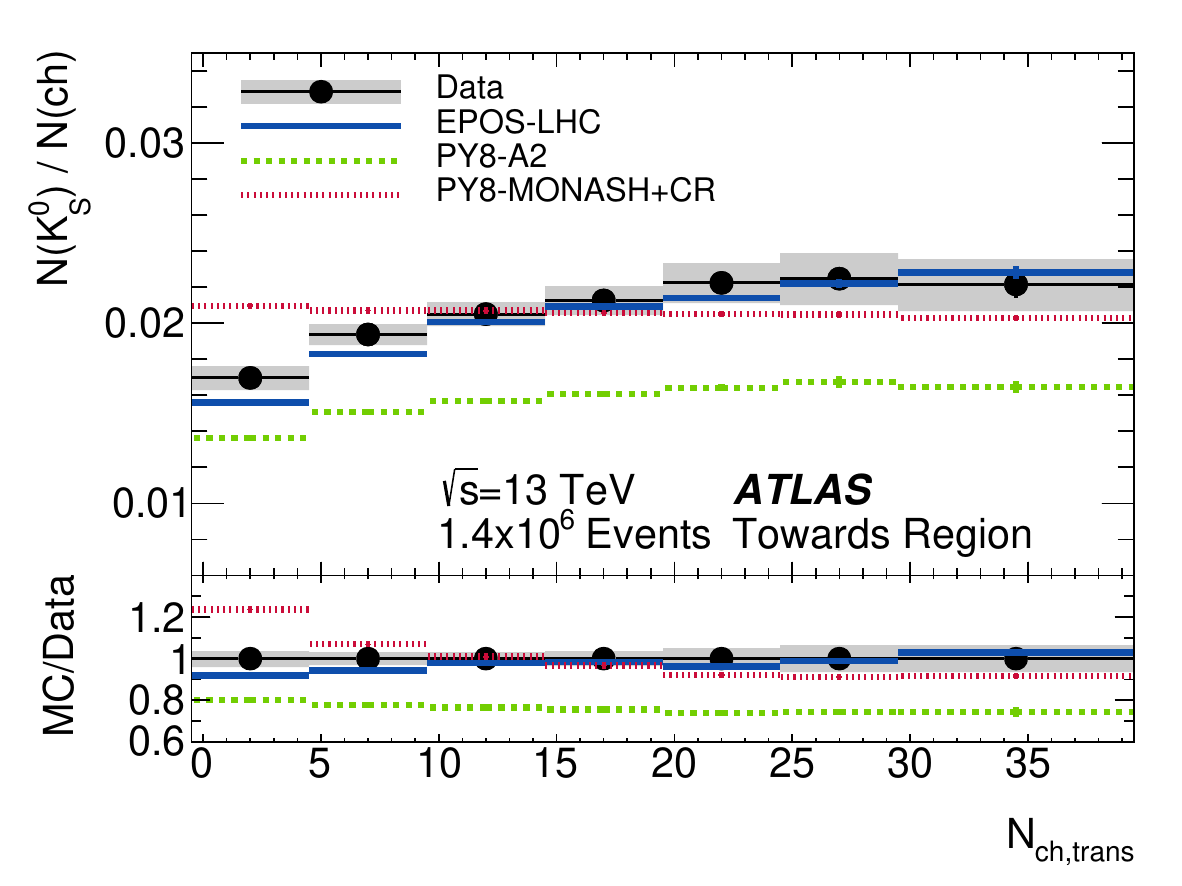}\label{fig:kch_n_to}}
\subfloat[]{\includegraphics[width=.48\linewidth]{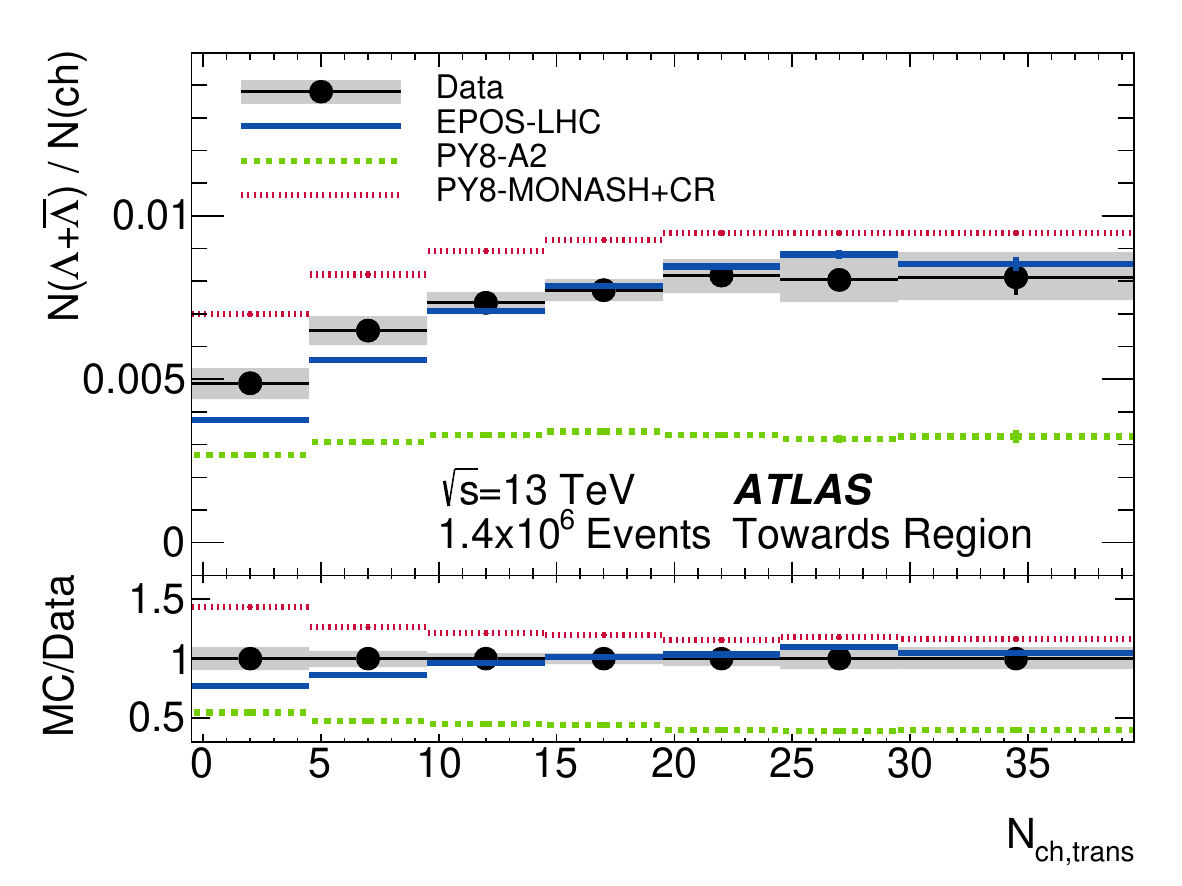}\label{fig:lch_n_to}}
\newline
\subfloat[]{
\includegraphics[width=0.48\textwidth]{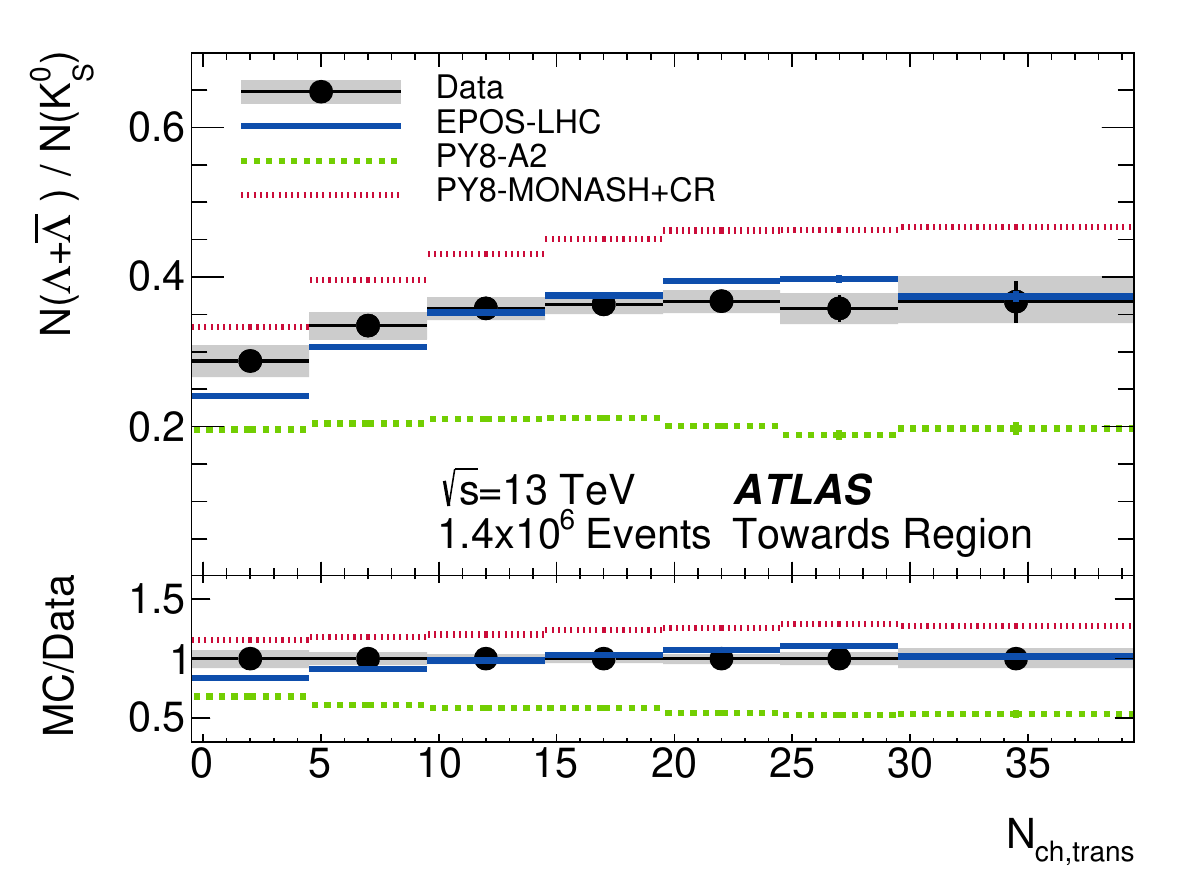}
\label{fig:lk_n_to}
}
\subfloat[]{
\includegraphics[width=0.48\textwidth]{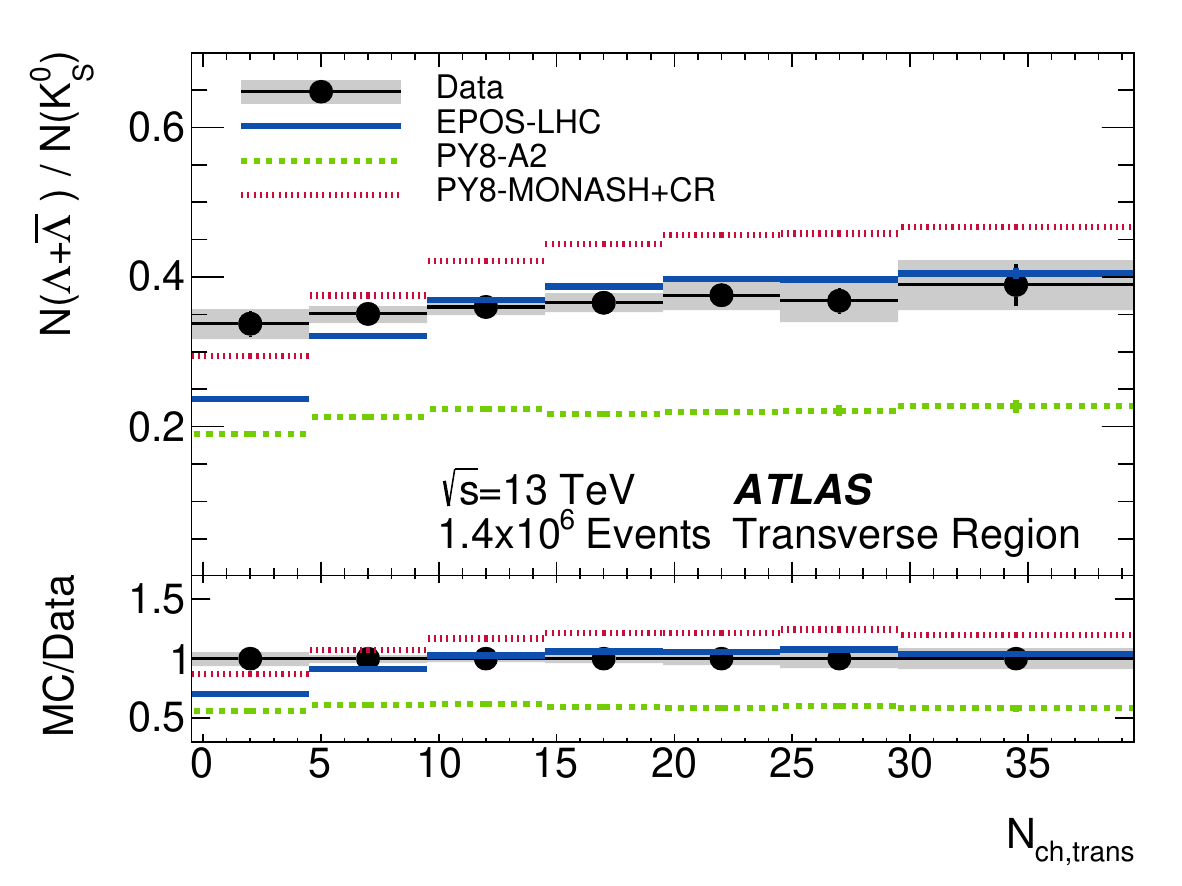}
\label{fig:lk_n_tr}
}
\caption{Comparison between data and MC simulation for several multiplicity ratios as a function of \nchtrans in events with leading jet \restrictedRange. Shown are the prompt charged-particle normalised (a) \Kshort and (b) \LLbar multiplicity yields in the towards region, and relative yields of \LLbar to \Kshort in the (c) towards and (d) transverse regions. Error bars show the statistical error and the shaded bands show the combination of statistical and systematic uncertainties.}
\label{fig:nch_ratio}
\end{figure}

The data discussed so far as being in the hard regime are explored further in Figure~\ref{fig:nch_ratio}. The sample of events whose leading jet lies inside \restrictedRange are included in these figures, with the per-event number of prompt charged-particles in the transverse region $\left(\nchtrans\right)$ used to define the plot abscissa. \nchtrans is used as a proxy for the number of soft and semi-hard interactions within the $pp$ collision in this subset of events.

The \Kshort and \LLbar yields as normalised to the charged-particle yield are shown in Figures~\ref{fig:kch_n_to} and \ref{fig:lch_n_to} in the towards region. The model disagreement is worse at small values of \nchtrans which correspond to events with little activity in the transverse region. EPOS-LHC performs the best here, with the \Pythia8 A2 model continuing to underestimate the yield while better modelling the shape. The \Pythia8 Monash+CR model predicts no dependence on the relative \Kshort yield with \nchtrans in the towards region, which is not in agreement with the data.

In Figures~\ref{fig:lk_n_to} and \ref{fig:lk_n_tr} the ratio of \LLbar to \Kshort multiplicity is presented in the towards and transverse regions.
The data show only a small suppression of the \LLbar to \Kshort ratio in the towards region for low values of \nchtrans, with the dependence on this ratio in the transverse region as a function of \nchtrans being even weaker still. Of the considered models, only the more simplistic (with regards to the modelling of strange production in the underlying-event) \Pythia8 A2 model correctly predicts that the \LLbar to \Kshort ratios are largely insensitive to the \nchtrans activity levels in the transverse region of the event. However, the absolute value of the ratio from the \Pythia8 A2 model is under predicted by around 40\%, due to the underproduction of \LLbar hadrons in the A2 model relative to data.

These trends may be contrasted with the ALICE measurement in Ref.~\cite{ALICE:2016fzo}. A similar enhancement of the strange yield relative to prompt charged-particles is observed by ALICE with respect to prompt charged pions for \Kshort and \LLbar.
This is seen here in Figures~\ref{fig:kch_n_to}, \ref{fig:lch_n_to} in the towards region of events at higher values of \nchtrans. Here the relative abundance of the strange hadrons is computed in a kinematic space which integrates over all values of leading-jet \pt above 10 \GeV in order to focus on the scaling of this ratio with the amount of activity in the underlying event, this allows for a more direct probe of colour reconnection effects. This is the opposite scaling trend than what was observed above in Figures~\ref{fig:kc_n_to} and \ref{fig:lc_n_to}, where the equivalent strange-particle yields in the towards regions were decreasing as a function of increasing leading-jet \pt above 10 \GeV. In these distributions the relative abundance of the strange hadrons is integrated over events with small and large levels of underlying event activity and the scaling is more strongly influenced by particle production from jet fragmentation and hadronisation processes at increasingly large momentum scales.
This highlights the importance of considering both of the choices of abscissa when investigating the modelling of colour-reconnection effects.

\section{Conclusion}
\label{sec:conclusion}

Properties of the underlying-event are investigated via the strange hadrons \Kshort, \Lam and \Lbar in ATLAS minimum-bias $pp$ collision data at $\sqrt{s} = 13~\TeV$. The hadrons are reconstructed via the identification of the displaced two-particle vertices corresponding to the decay modes $K_{S}^{0}\rightarrow\pi^+\pi^-$, $\Lambda\rightarrow\pi^-p$ and $\bar\Lambda\rightarrow\pi^+\bar{p}$.

\Kshort, \Lam and \Lbar multiplicity ratios are constructed normalised to the number of events, or to the prompt charged-particle multiplicity in the underlying event `toward', `transverse' and `away' regions relative to the azimuthal angle $\phi$ of the leading charged-particle jet. They are compared to different MC simulation models.

\Pythia8 A2 is able to describe the dependence of these observables on the leading-jet \pt for both of the \Kshort and \LLbar distributions, but underestimates the yields by around 40\%--50\%. \Pythia8 Monash+CR displays an enhanced \Kshort yield and a significantly enhanced \LLbar yield bringing it more in alignment with data. EPOS-LHC is better able to model the rise with leading-jet \pt, however it typically plateaus out too early -- and even decreases in some event-normalised observables at yet higher leading-jet \pt, which is in contrast to what was observed in the data.

A further selection considers events whose leading jet has \restrictedRange.
The $\Lam/\Kshort$ ratio is observed to have little dependence of the number of prompt charged-particles in the transverse region, which acts here as a proxy for the number of multi-parton interactions in the event. An enhancement of the \Kshort and \LLbar to prompt charged-particles ratios in the towards region is observed for larger values of the number of prompt charged-particles in the transverse region. The EPOS-LHC model is in best agreement with these data.

These data measurements are particularly sensitive to the theoretical modelling and MC tuning of strange and baryon production during hadronisation and may be used to constrain theoretical modelling of non-perturbative effects within individual $pp$ interactions.

\section*{Acknowledgements}
%

%
%

%
%

We thank CERN for the very successful operation of the LHC and its injectors, as well as the support staff at
CERN and at our institutions worldwide without whom ATLAS could not be operated efficiently.

The crucial computing support from all WLCG partners is acknowledged gratefully, in particular from CERN, the ATLAS Tier-1 facilities at TRIUMF/SFU (Canada), NDGF (Denmark, Norway, Sweden), CC-IN2P3 (France), KIT/GridKA (Germany), INFN-CNAF (Italy), NL-T1 (Netherlands), PIC (Spain), RAL (UK) and BNL (USA), the Tier-2 facilities worldwide and large non-WLCG resource providers. Major contributors of computing resources are listed in Ref.~\cite{ATL-SOFT-PUB-2023-001}.

We gratefully acknowledge the support of ANPCyT, Argentina; YerPhI, Armenia; ARC, Australia; BMWFW and FWF, Austria; ANAS, Azerbaijan; CNPq and FAPESP, Brazil; NSERC, NRC and CFI, Canada; CERN; ANID, Chile; CAS, MOST and NSFC, China; Minciencias, Colombia; MEYS CR, Czech Republic; DNRF and DNSRC, Denmark; IN2P3-CNRS and CEA-DRF/IRFU, France; SRNSFG, Georgia; BMBF, HGF and MPG, Germany; GSRI, Greece; RGC and Hong Kong SAR, China; ISF and Benoziyo Center, Israel; INFN, Italy; MEXT and JSPS, Japan; CNRST, Morocco; NWO, Netherlands; RCN, Norway; MNiSW, Poland; FCT, Portugal; MNE/IFA, Romania; MESTD, Serbia; MSSR, Slovakia; ARIS and MVZI, Slovenia; DSI/NRF, South Africa; MICIU/AEI, Spain; SRC and Wallenberg Foundation, Sweden; SERI, SNSF and Cantons of Bern and Geneva, Switzerland; NSTC, Taipei; TENMAK, T\"urkiye; STFC/UKRI, United Kingdom; DOE and NSF, United States of America.

Individual groups and members have received support from BCKDF, CANARIE, CRC and DRAC, Canada; PRIMUS 21/SCI/017, CERN-CZ and FORTE, Czech Republic; COST, ERC, ERDF, Horizon 2020, ICSC-NextGenerationEU and Marie Sk{\l}odowska-Curie Actions, European Union; Investissements d'Avenir Labex, Investissements d'Avenir Idex and ANR, France; DFG and AvH Foundation, Germany; Herakleitos, Thales and Aristeia programmes co-financed by EU-ESF and the Greek NSRF, Greece; BSF-NSF and MINERVA, Israel; Norwegian Financial Mechanism 2014-2021, Norway; NCN and NAWA, Poland; La Caixa Banking Foundation, CERCA Programme Generalitat de Catalunya and PROMETEO and GenT Programmes Generalitat Valenciana, Spain; G\"{o}ran Gustafssons Stiftelse, Sweden; The Royal Society and Leverhulme Trust, United Kingdom.

In addition, individual members wish to acknowledge support from CERN: European Organization for Nuclear Research (CERN PJAS); Chile: Agencia Nacional de Investigaci\'on y Desarrollo (FONDECYT 1190886, FONDECYT 1210400, FONDECYT 1230812, FONDECYT 1230987); China: Chinese Ministry of Science and Technology (MOST-2023YFA1605700), National Natural Science Foundation of China (NSFC - 12175119, NSFC 12275265, NSFC-12075060); Czech Republic: Czech Science Foundation (GACR - 24-11373S), Ministry of Education Youth and Sports (FORTE CZ.02.01.01/00/22\_008/0004632), PRIMUS Research Programme (PRIMUS/21/SCI/017); EU: H2020 European Research Council (ERC - 101002463); European Union: European Research Council (ERC - 948254, ERC 101089007), Horizon 2020 Framework Programme (MUCCA - CHIST-ERA-19-XAI-00), European Union, Future Artificial Intelligence Research (FAIR-NextGenerationEU PE00000013), Italian Center for High Performance Computing, Big Data and Quantum Computing (ICSC, NextGenerationEU); France: Agence Nationale de la Recherche (ANR-20-CE31-0013, ANR-21-CE31-0013, ANR-21-CE31-0022, ANR-22-EDIR-0002), Investissements d'Avenir Labex (ANR-11-LABX-0012); Germany: Baden-Württemberg Stiftung (BW Stiftung-Postdoc Eliteprogramme), Deutsche Forschungsgemeinschaft (DFG - 469666862, DFG - CR 312/5-2); Italy: Istituto Nazionale di Fisica Nucleare (ICSC, NextGenerationEU), Ministero dell'Università e della Ricerca (PRIN - 20223N7F8K - PNRR M4.C2.1.1); Japan: Japan Society for the Promotion of Science (JSPS KAKENHI JP21H05085, JSPS KAKENHI JP22H01227, JSPS KAKENHI JP22H04944, JSPS KAKENHI JP22KK0227); Netherlands: Netherlands Organisation for Scientific Research (NWO Veni 2020 - VI.Veni.202.179); Norway: Research Council of Norway (RCN-314472); Poland: Polish National Agency for Academic Exchange (PPN/PPO/2020/1/00002/U/00001), Polish National Science Centre (NCN 2021/42/E/ST2/00350, NCN OPUS nr 2022/47/B/ST2/03059, NCN UMO-2019/34/E/ST2/00393, UMO-2020/37/B/ST2/01043, UMO-2021/40/C/ST2/00187, UMO-2022/47/O/ST2/00148, UMO-2023/49/B/ST2/04085); Slovenia: Slovenian Research Agency (ARIS grant J1-3010); Spain: Generalitat Valenciana (Artemisa, FEDER, IDIFEDER/2018/048), Ministry of Science and Innovation (MCIN \& NextGenEU PCI2022-135018-2, MICIN \& FEDER PID2021-125273NB, RYC2019-028510-I, RYC2020-030254-I, RYC2021-031273-I, RYC2022-038164-I), PROMETEO and GenT Programmes Generalitat Valenciana (CIDEGENT/2019/023, CIDEGENT/2019/027); Sweden: Swedish Research Council (Swedish Research Council 2023-04654, VR 2018-00482, VR 2022-03845, VR 2022-04683, VR 2023-03403, VR grant 2021-03651), Knut and Alice Wallenberg Foundation (KAW 2018.0157, KAW 2018.0458, KAW 2019.0447, KAW 2022.0358); Switzerland: Swiss National Science Foundation (SNSF - PCEFP2\_194658); United Kingdom: Leverhulme Trust (Leverhulme Trust RPG-2020-004), Royal Society (NIF-R1-231091); United States of America: U.S. Department of Energy (ECA DE-AC02-76SF00515), Neubauer Family Foundation.

%
%


%
%
%
%
%
%

%
%
%
%
%
%

%
%

%
%
%
\printbibliography
\clearpage
 
\begin{flushleft}
\hypersetup{urlcolor=black}
{\Large The ATLAS Collaboration}

\bigskip

\AtlasOrcid[0000-0002-6665-4934]{G.~Aad}$^\textrm{\scriptsize 103}$,
\AtlasOrcid[0000-0001-7616-1554]{E.~Aakvaag}$^\textrm{\scriptsize 16}$,
\AtlasOrcid[0000-0002-5888-2734]{B.~Abbott}$^\textrm{\scriptsize 121}$,
\AtlasOrcid[0000-0002-0287-5869]{S.~Abdelhameed}$^\textrm{\scriptsize 117a}$,
\AtlasOrcid[0000-0002-1002-1652]{K.~Abeling}$^\textrm{\scriptsize 55}$,
\AtlasOrcid[0000-0001-5763-2760]{N.J.~Abicht}$^\textrm{\scriptsize 49}$,
\AtlasOrcid[0000-0002-8496-9294]{S.H.~Abidi}$^\textrm{\scriptsize 29}$,
\AtlasOrcid[0009-0003-6578-220X]{M.~Aboelela}$^\textrm{\scriptsize 44}$,
\AtlasOrcid[0000-0002-9987-2292]{A.~Aboulhorma}$^\textrm{\scriptsize 35e}$,
\AtlasOrcid[0000-0001-5329-6640]{H.~Abramowicz}$^\textrm{\scriptsize 152}$
% ...
\bigskip
\\

% ...
$^{16}$Department for Physics and Technology, University of Bergen, Bergen; Norway.\\
% ...
$^{29}$Physics Department, Brookhaven National Laboratory, Upton NY; United States of America.\\
% ...
$^{35}$$^{(a)}$Facult\'e des Sciences Ain Chock, Universit\'e Hassan II de Casablanca;$^{(b)}$Facult\'{e} des Sciences, Universit\'{e} Ibn-Tofail, K\'{e}nitra;$^{(c)}$Facult\'e des Sciences Semlalia, Universit\'e Cadi Ayyad, LPHEA-Marrakech;$^{(d)}$LPMR, Facult\'e des Sciences, Universit\'e Mohamed Premier, Oujda;$^{(e)}$Facult\'e des sciences, Universit\'e Mohammed V, Rabat;$^{(f)}$Institute of Applied Physics, Mohammed VI Polytechnic University, Ben Guerir; Morocco.\\
% ...
$^{44}$Physics Department, Southern Methodist University, Dallas TX; United States of America.\\
% ...
$^{49}$Fakult\"{a}t Physik , Technische Universit{\"a}t Dortmund, Dortmund; Germany.\\
% ...
$^{55}$II. Physikalisches Institut, Georg-August-Universit\"{a}t G\"ottingen, G\"ottingen; Germany.\\
% ...
$^{103}$CPPM, Aix-Marseille Universit\'e, CNRS/IN2P3, Marseille; France.\\
% ...
$^{117}$$^{(a)}$New York University Abu Dhabi, Abu Dhabi;$^{(b)}$United Arab Emirates University, Al Ain; United Arab Emirates.\\
% ...
$^{121}$Homer L. Dodge Department of Physics and Astronomy, University of Oklahoma, Norman OK; United States of America.\\
% ...
$^{152}$Raymond and Beverly Sackler School of Physics and Astronomy, Tel Aviv University, Tel Aviv; Israel.\\
% ...
$^{a}$ Also Affiliated with an institute covered by a cooperation agreement with CERN.\\
$^{b}$ Also at An-Najah National University, Nablus; Palestine.\\
$^{c}$ Also at Borough of Manhattan Community College, City University of New York, New York NY; United States of America.\\
$^{d}$ Also at Center for Interdisciplinary Research and Innovation (CIRI-AUTH), Thessaloniki; Greece.\\
$^{e}$ Also at Centro Studi e Ricerche Enrico Fermi; Italy.\\
$^{f}$ Also at CERN, Geneva; Switzerland.\\
% ...

\end{flushleft}


%

%
%

%

%
%
%
%

\end{document}